\documentclass[aps,prb,twocolumn,showpacs,superscriptaddress,groupedaddress]{revtex4}

\usepackage{graphicx}
\usepackage{float}
\usepackage{sidecap}
\usepackage{dcolumn}
\usepackage{bm}
\usepackage{amssymb}
\usepackage{amsmath}	
\usepackage{color}
\usepackage{multirow}

\begin{document}


\widetext

\title{Atomistic theoretical study of electronic and polarization properties of \\ elliptical, single and vertically stacked InAs quantum dots}


\author{Muhammad Usman} \email{usman@alumni.purdue.edu} \affiliation{Tyndall National Institute, Lee Maltings, Dyke Parade, Cork, Ireland.} 
\vskip 0.25cm



\begin{abstract}
The demonstration of isotropic polarization response from semiconductor quantum dots (QDs) is a crucial step towards the design of several optoelectronic technologies. Among many parameters that impact the degree of polarization (DOP$_{[\overrightarrow{n}]}$) of a QD system, the shape asymmetry is a critical factor. We perform multi-million-atom simulations to study the impact of the elliptical shapes on the electronic and polarization properties of single and vertically stacked InAs QDs. The comparison between a low aspect ratio (AR) and a high AR QD reveals that the electronic and the polarization properties strongly depend on the AR of the QD; the elongation of a tall QD allows tuning of the DOP$_{[\overrightarrow{n}]}$ over a much wider range. We then extend our analysis to an experimentally reported vertical stack of nine QDs (9-VSQDs) that has shown significant potential to achieve isotropic polarization properties. We analyse the contribution from the shape asymmetry in the experimentally measured large in-plane polarization anisotropy. Our analysis shows that the orientation of the base elongation controls the sign of the DOP$_{[\overrightarrow{n}]}$; however the magnitude of the base elongation has only a very little impact on the magnitude of the DOP$_{[\overrightarrow{n}]}$. We further predict that the elliptical shape of the 9-VSQDs can only tune either DOP$_{[110]}$ or DOP$_{[\overline{1}10]}$ for the isotropic response. Our model results, in agreement with the TEM findings, suggest that the experimentally grown 9-VSQDs has either a circular-base or a slightly [$\overline{1}$10] elongated base. Overall the detailed investigation of the DOP$_{[\overrightarrow{n}]}$ as a function of the QD shape asymmetry provides a theoretical guidance for the continuing experimental efforts to achieve tailored polarization properties from the QD nano-structures for the design of optical devices.               
\end{abstract}

\maketitle


\section{Introduction}

Deployment of semiconductor quantum dots (QDs) in the active region of optical devices offers unique electronic and optical properties which can be exploited to design several optoelectronic technologies ranging from lasers\cite{Salhi_1} to semiconductor optical amplifiers (SOAs)\cite{Akiyama_1} or single photon sources\cite{Dousse_1}, where they have successfully overcome critical challenges such as extremely low threshold, high speed response, or entangled photon emission, respectively. However, in these applications, a critical design parameter is the polarization response of QDs, typically characterized in terms of either degree of polarization [DOP = (TE-TM)/(TE+TM)]\cite{Usman_1, Usman_2} or TM/TE ratio\cite{Fortunato_1, Usman_6}, where TE-mode is measured along a direction in the plane of the QD, and TM-mode is measured along the growth [001] direction for the GaAs(001) QDs. Engineering of QD nanostrcutures to achieve isotropic polarization (DOP $\sim$ 0) is critical for the implementation of several optoelectronic devices, for example semiconductor optical amplifiers (SOA's).    

InAs QDs grown by the Stranski-Krastonov (SK) self-assembly growth process typically exhibit very poor polarization response (DOP close to 1.0) due to the large compressive biaxial strain surrounding the flat shapes of the QDs. The strain induced splitting between the heavy hole (HH) and the light hole (LH) valence bands leads to a dominant HH character in the few top most valence band states, thus significantly suppressing the TM-mode. Therefore, previous studies of the single InAs QDs have reported very high values of the DOP, typically larger than 0.8.\cite{Usman_1, Fortunato_1, Saito_1, Inoue_1} 

The polarization response of InAs QDs is influenced by several parameters such as crystal/atomic symmetry, QD shape, composition profile etc. The atomistic asymmetry of the underlying zincblende crystals implies that the [110] and [$\overline{1}$10] directions are inequivalent. This lowers the overall symmetry of a perfectly circular dome-shaped QD from C$_{\infty v}$ to C$_{2v}$. As a result, TE-mode in the plane of the QD does not remain symmetric and significant in-plane anisotropy may be observed even for an ideal circular-base InAs QD.\cite{Usman_5} Therefore, a single value of the DOP is not sufficient to characterize the polarization response of the QD systems. This, in the past studies\cite{Usman_1, Usman_5, Usman_6}, has lead us to define a direction-dependent value of the DOP,

\begin{equation}
  \label{eq:dop}
   \begin{array}{cc}
   DOP_{[\overrightarrow{n}]} = \frac{(TE_{[\overrightarrow{n}]}-TM_{[001]})}{(TE_{[\overrightarrow{n}]}+TM_{[001]})}    \end{array}
\end{equation}
\\
where the direction, [$\overrightarrow{n}$] = [110] or [$\overline{1}$10], associated with the DOP$_{[\overrightarrow{n}]}$ is same as the direction of the TE$_{[\overrightarrow{n}]}$-mode in the plane of the QD.         

The value of the DOP$_{[\overrightarrow{n}]}$ also strongly depends on the shape of the QDs, which is significantly affected by the growth dynamics of the self-assembly process during the growth of the capping layers and the post-growth annealing processes\cite{Biasoil_1}. As a result, the shape of a SK self-assembled QD is far from being perfectly circular or square, as typically assumed in the past theoretical studies of the polarization properties. Several experimental investigations have suggested that the actual shape of the QDs significantly deviates from the ideal circular-base (for dome or lens) or square-base (for pyramid), and usually tends to elongate along the [110]\cite{Stevenson_1, Plumhof_1, Pryor_1, Fricke_1} or along the [$\overline{1}$10]\cite{Krapek_1, Hospodkova_1, Songmuang_1, Favero_1} directions.  

Furthermore, recent advancements in the growth techniques have allowed to control the shape of QDs, leading to the fabrication of strongly elongated QD like nano-structures.\cite{Dusanowski_1} These offer an enhanced exciton oscillator strength and allow the realization of single exciton-single photon coupling to build the fundamental blocks for the solid state quantum information.\cite{Favero_1} Such elongations of the QDs along the [$\overrightarrow{n}$] = [110] or [$\overline{1}$10] direction can significantly alter the value of the DOP$_{[\overrightarrow{n}]}$ and may be exploited to achieve tailored polarization response for a desired operation.  

\textbf{\textit{Brief overview of the past theoretical studies:}} Despite significant experimental evidence for the elongation of the QD shapes and its prospective potential to tune the polarization properties, the impact of the base elongations on the value of the DOP$_{[\overrightarrow{n}]}$ is only barely known. The previous theoretical investigations of the QD elongations are focused on the study and design of the fine structure splitting (FSS = energetic difference between the two bright excitons, e$_{[\overline{1}10]}$ - e$_{[110]}$)~\cite{Schliwa_1, Young_1, Krapek_1, Seguin_1, Plumhof_1, Singh_1, Singh_2} or the spin polarization~\cite{Pryor_1}, with very little emphasis given to the study of the polarization properties (comparison of the magnitudes of the TE and TM modes).\cite{Schliwa_1, Sheng_1, Mlinar_1} 

Sheng \textit{et al.}\cite{Sheng_1} applied an effective bond-orbital model and discussed the impact of the base elongations on the in-plane polarization anisotropy of the InGaAs QDs defined as:

\begin{equation}
  \label{eq:pol}
   \begin{array}{cc}
   Pol_{||} = \frac{(TE_{[\overline{1}10]}-TE_{[110]})}{(TE_{[\overline{1}10]}+TE_{[110]})}    \end{array}
\end{equation}
\\
and they concluded that the electron-electron interactions and the alloy intermixing have very little contributions in determining the polarization properties of the QDs. 

Mlinar \textit{et al.}\cite{Mlinar_1}, using an atomistic pseudo-potential model, focused on the InGaAs QDs and studied the impact of the (In,Ga)As alloy randomness on the polarization properties. They concluded that the alloy composition fluctuations can significantly change the in-plane polarization anisotropy and therefore the experimentally measured polarization anisotropy may not be considered as a reliable measure of the QD shape asymmetry. 

Schliwa \textit{et al.}~\cite{Schliwa_1}, based on their \textbf{k}$\centerdot$\textbf{p} calculations, studied the impact of piezoelectricity on the QD optical properties. Although they provided a detailed investigation of the electronic and optical properties of the dome and pyramidal shaped QDs as a function of their vertical aspect ratio (height/base), only one case of the pyramidal shaped QDs (series D in their paper) was investigated for the study of the base elongations (lateral aspect ratio). Furthermore, in their study of the lateral aspect ratios of the pyramidal QDs, they kept the overall volume of the QDs unchanged by altering both the heights and the base diameters of the QDs. Since the QD energies are strongly influenced by a change in their height parameter\cite{Usman_3}, so the reported results did not isolate the impact of the QD base elongations. 

Nevertheless, a comprehensive quantitative analysis of the impact of the QD base elongations on the polarization dependent room temperature ground state optical emissions still remain unavailable. This paper, therefore, aims to bridge this gap by providing a detailed study of the DOP$_{[\overrightarrow{n}]}$ and Pol$_{||}$ for the [110]- and [$\overline{1}$10]-elongated QDs. For the calculation of the polarization dependent optical modes (TE and TM), we take into account the highest five valence band states, instead of just a single top most valence band state, in accordance with the recent studies~\cite{Usman_2, Usman_5} where it has been shown that the calculation of the room temperature ground state optical spectra must involve multiple closely spaced valence band states to accurately model the in-plane polarizability and to avoid discrepancy between the theory and experiments. Our calculations show that despite tuning of the DOP$_{[\overrightarrow{n}]}$ over a wide range, the elliptical shapes of the single QDs do not lead to an isotropic polarization. Therefore, we extend our study to multi-layer QD stacks. 
 
\textbf{\textit{Isotropic polarization from multi-layer stacks:}} Past theoretical\cite{Usman_1, Saito_1} and experimental\cite{Inoue_1, Alonso_1, Ikeuchi_1} studies have shown that the polarization response of the QDs can be drastically improved by growing large vertical stacks of QDs (VSQDs), consisting of many closely spaced QD layers to exploit inter-dot strain and electronic couplings. A recent experimental study~\cite{Inoue_1} has shown that a vertical stack of nine QDs (9-VSQDs) exhibits DOP$_{[110]}$ = -0.6; however the measured PL spectra showed a large anisotropy in the in-plane polarization modes: TE$_{[\overline{1}10]} \gg$ TE$_{[110]}$. our atomistic calculations~\cite{Usman_1}, based on an assumption of ideal circular-base dome-shape for the QD layers, reported TE$_{[\overline{1}10]} \gg$ TE$_{[110]}$ in agreement with experimental PL spectra. We attributed this large in-plane polarization anisotropy to a small increase in the TM$_{[001]}$-mode (coming from an enhanced HH/LH intermixing) and a large decrease in the TE$_{[110]}$-mode due to the [$\overline{1}$10]-oriented hole wave function confinements. 

A good qualitative agreement of our calculations with the experimental results assuming ideal circular-base for the QDs leads to a fundamental question that how much is the contribution from the QD shape asymmetry in the polarization anisotropy for the 9-VSQDs? This work, based on multi-million-atom calculations, provides the answer that the interfacial hole wave function confinements have major contribution in the experimentally measured in-plane polarization anisotropy, which is counter-intuitive to the common notion where the shape asymmetry is considered mainly response for such anisotropies.\cite{Humlicek_1, Alonso_1}  

Furthermore, as an [110]-elongation increases DOP$_{[110]}$ and reduces DOP$_{[\overline{1}10]}$, so we investigate the possibility to exploit it to balance the built-in in-plane anisotropy such that to achieve DOP$_{[110]} \sim$ 0 and DOP$_{[\overline{1}10]} \sim$ 0? This would lead to the design of the QD based SOAs independent of the in-plane direction. Our study reveals an interesting property of the 9-VSQDs that its polarization response is very sensitive to the orientation of its elongation and both TE-modes can not be reduced below TM-mode. Therefore, either DOP$_{[\overline{1}10]}$ or DOP$_{[110]}$ can be tuned for an isotropic polarization, and not both of them simultaneously. 

Finally, the quantitative study of the elongation dependent DOP$_{[\overrightarrow{n}]}$ also helps us to determine the geometry of the 9-VSQDs. Our theoretical calculations accurately predict [$\overline{1}$10] elongation of the 9-VSQDs in the experiment, which is also consistent with the TEM images\cite{Kita_1}.     

The remainder of the paper is organized in the following sections: section II defines QD geometry parameters and describes the three types of elliptical shapes that we study in this paper. Section III documents our methodologies. Our results for two different AR single QDs are presented in sections IV-A and IV-B. Section IV-C is about the vertical stack of nine QD layers (9-VSQDs). The sections IV-A, IV-B, and IV-C are written as self-contained sections so that a reader interested in only one type of QDs may only require to read the corresponding section. Finally, we provide an overall summary and main conclusions of our results in section V.                       


\section{Simulated Quantum Dot Systems}

\subsection{Geometry Parameters} 

\begin{figure}
\includegraphics[scale=0.37]{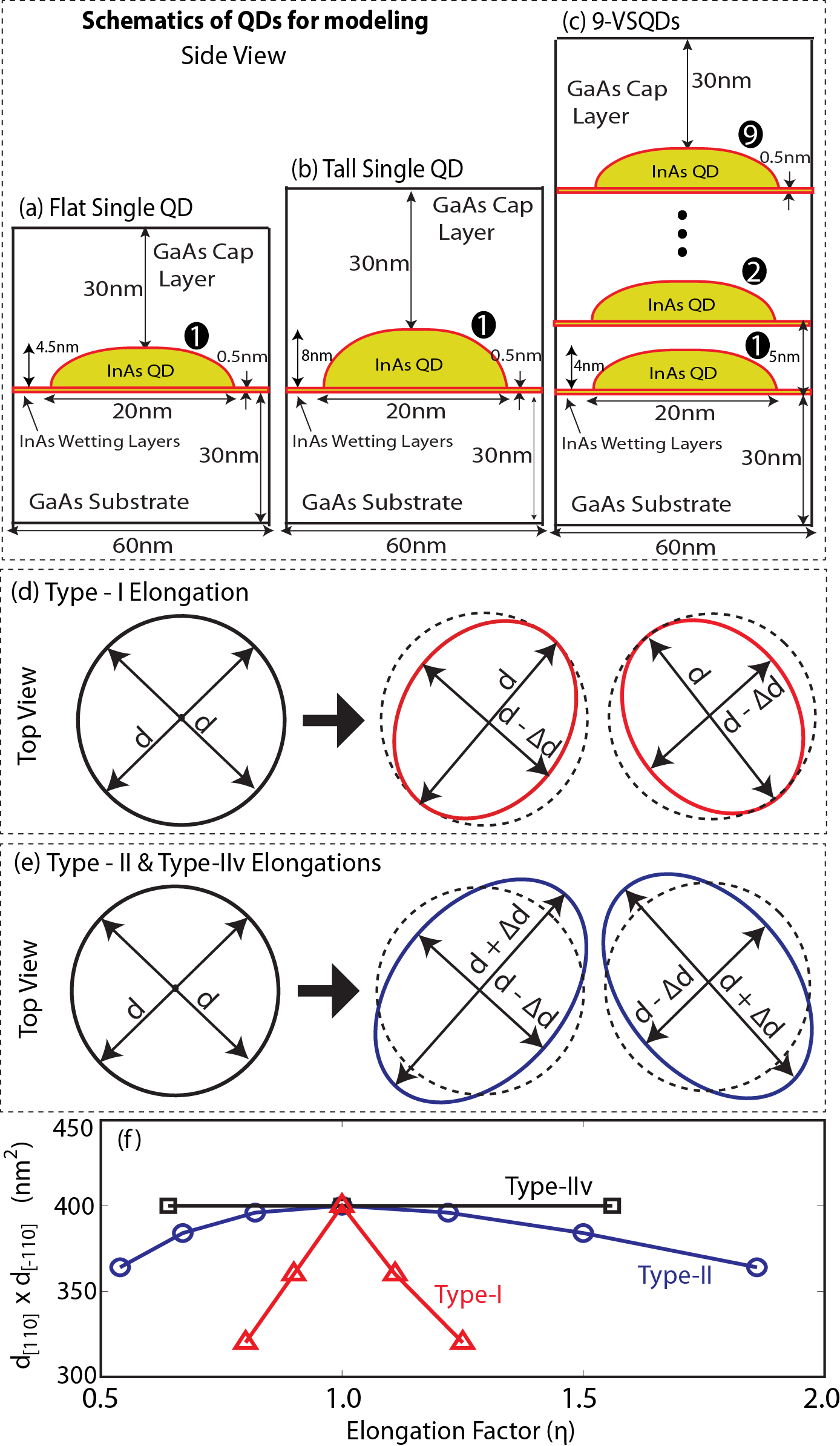}
\caption{Schematic of quantum dots are shown for theoretical modeling. (a) A low aspect ratio (flat) dome-shaped InAs quantum dot with the base diameter and height of 20 nm and 4.5 nm, respectively. (b) A high aspect ratio (tall) dome-shaped InAs quantum dot with the base diameter and height of 20 nm and 8 nm, respectively. (c) A vertical stack consisting of nine InAs QDs (9-VSQDs), each with the base diameter and height of 20 nm and 4 nm, respectively. The geometry parameters are directly taken from the experimental study\cite{Inoue_1}. (d) Top view illustrating the base elongations for the Type-I elongation. The elliptical shape is formed by decreasing the diameter either along the [110] or along the [-110] direction. (e) Top view illustrating the base elongations for the Type-II and Type-IIv elongations. The elliptical shape is formed by simultaneously decreasing (increasing) the base diameter along the [110] direction and increasing (decreasing) the base diameter along the [-110] direction. For the Type-IIv elongation, we select values for the d$_{[110]}$ and d$_{[\overline{1}10]}$ such that to keep the overall volume of the QD fixed. (f) Plots of the products of the QD diameters along the [110] and the [$\overline{1}$10] directions, d$_{[110]} \times$d$_{[\overline{1}10]}$, as a function of the elongation factor ($\eta$). For the fixed QD height, this product is directly proportional to the volume of the QD.}
\label{fig:Fig1}
\end{figure}

Figs.\ref{fig:Fig1}(a), (b), and (c) show three quantum dot geometries simulated in this study. The InAs quantum dots are embedded inside large GaAs buffers comprised of $\approx$ 15 million atoms (60$\times$60$\times$60 nm$^3$) for (a) and (b), and $\approx$ 25 million atoms (60$\times$60$\times$106 nm$^3$) for (c). The quantum dots are placed on top of 0.5 nm thick InAs wetting layers. 

We study three dome-shaped quantum dot systems: (i) a low aspect ratio (AR) InAs QD with 20 nm diameter (d) and 4.5 nm height (h), (AR = h/d = 0.225); (ii) a high AR InAs QD with d = 20 nm and h = 8.0 nm (AR = h/d = 0.40); (iii) a vertical stack comprised of nine QD layers (9-VSQDs) separated by 5 nm thick GaAs spacer layers, where each layer consists of an InAs QD with d = 20 nm and h = 4 nm. The geometrical parameters of this 9-VSQDs are taken from the recent experimental\cite{Inoue_1, Ikeuchi_1} and theoretical~\cite{Usman_1} studies where it has shown great technological relevance for achieving isotropic polarization response. 

In remainder of this paper, we label the single QD with AR = 0.225 as a "flat" QD and the single QD with AR = 0.40 as a "tall" QD. In a previous study~\cite{Usman_5}, it has been shown that the flat and tall QDs with an ideal circular-base exhibit drastically different electronic and polarization properties. The hole wave functions tend to reside in the HH pockets for the tall QDs, when the AR $\gtrsim$ 0.25. This introduces a large anisotropy in the in-plane polarization (Pol$_{||}$). Therefore, this paper analyses the impact of the base elongations for both types of the QDs. Furthermore, as the 9-VSQDs consists of strongly coupled QDs so it can essentially be considered as an extension of the single tall QD with a very large AR $\cong$ 45/20 = 2.25. Our calculations presented in the section (IV) confirm that the two single QDs (flat and tall) exhibit drastically different polarization properties as a function of their base elongations, and the 9-VSQDs overall exhibiting many similar characteristics as that of the tall QD. 

Note that a typical SK growth of a large vertical QD stack generally results in an increase in the size of the upper layer QDs\cite{Xie_1}, however, no such increase in the QD dimensions is reported in the experimental study\cite{Inoue_1}. Therefore, we keep the size of the QDs uniform for the 9-VSQDs.  

\vspace{1mm}


\subsection{Elongation along High Symmetry Axis} 

In order to study the impact of the base elongations of the three QD systems described above along the high symmetry crystallographic directions ([110] and [$\overline{1}$10]) on their electronic and polarization properties, we consider three types of the elliptical shapes, defined by an elongation factor $\eta$ = d$_{[110]}$/d$_{[\overline{1}10]}$ which is a ratio of the QD lateral diameters along the [110] and [$\overline{1}$10] directions, for the above mentioned three QD systems: \\ \\ Type-I: As schematically shown in Fig.~\ref{fig:Fig1}(d), the diameter d$_{[110]}$ (d$_{[\overline{1}10]}$) is reduced by $\bigtriangleup$d for [110] ([$\overline{1}$10]) elongation, while keeping the other diameter d$_{[\overline{1}10]}$ (d$_{[110]}$) fixed at 20 nm. This will reduce the overall volume of the QD. \\ \\ Type-II: For this type of elongation, as schematically shown in the Fig.~\ref{fig:Fig1}(e), we simultaneously change both d$_{[110]}$  and d$_{[\overline{1}10]}$ diameters by equal amounts $\bigtriangleup$d (one diameter is increased by $\bigtriangleup$d and the other diameter is decreased by $\bigtriangleup$d). Once again, the volume of the QD decreases, however, at much slower rate compared to the Type-I case. \\ \\ Type-IIv: As a special case of the Type-II elongation, we again change the QD diameters along the [110] and [$\overline{1}$10] directions similar to the Type-II case, but keep the overall QD volume unchanged. Most of the previous theoretical studies\cite{Schliwa_1, Singh_1, Singh_2, Pryor_1, Sheng_1, Mlinar_1} have only analysed this type of elongation, so it allows us to make a direct comparison with the existing results. For a circular-base dome-shape QD with d = 20 nm and h = 4.5 nm, the volume of the QD is (1/3)$\pi$d$^2$h $\propto$ d$^2$ $\propto$ 20$^2$. Now to keep the volume unchanged, the other reasonable choice for the diameters d$_{[110]}$ and d$_{[\overline{1}10]}$ is 25 nm and 16 nm that results in d$_{[110]} \times$d$_{[\overline{1}10]}$ = 400. Therefore for the Type-IIv elongation, we choose between 25 nm and 16 nm for the diameters d$_{[110]}$ and d$_{[\overline{1}10]}$.

As the volume of an ellipsoidal QD is proportional to the product of diameters along its major and minor axes (d$_{[110]}$ and d$_{[\overline{1}10]}$), we plot this product as a function of the elongation factor $\eta$ for the three types of elongations in Fig.~\ref{fig:Fig1}(f), quantitatively showing the decrease in the QD volume for the three types of elongations. As it will become clear in the later sections, the large decrease in the QD volume ($\bigtriangleup$V) for the Type-I elongation will significantly impact the electronic and the optical properties, even dominating impact of the diameter changes. For the Type-II elongation, relatively much smaller decrease in the volume will compete with the changes in the diameter and the net impact will be from $\bigtriangleup$d for the small values of $\eta$, and from the $\bigtriangleup$V for the large values of $\eta$.     

It should be noted that in all of the three types of elongations, we keep the height of the QDs fixed. It has been shown\cite{Usman_3} that the electronic and optical properties are very sensitive to the height of the QDs, so by keeping it unchanged, we eliminate its contribution and only focus on the impact of the base elongations. We also specify that the previous theoretical investigations\cite{Schliwa_1, Sheng_1} of the QD elongations have used the term "lateral aspect ratio" for the ratio of the QD base diameters, which is equivalent to the elongation factor $\eta$ defined in this study. In order to avoid confusion, we use aspect ratio for the ratio of the QD height and base diameter; and use elongation factor for the ratio of the base diameters. Finally, by definition, the elongation factor $\eta$ is = 1.0, $>$1.0, and $<$1.0 for the circular, [110]-elongated, and $[\overline{1}10]$-elongated QDs, respectively.    


\section{Methodologies}

The atomistic simulations are performed using NEMO 3-D simulator\cite{Klimeck_1, Klimeck_2, Klimeck_3}, which is based on strain energy minimization by using the valence force field (VFF) model\cite{Keating_1, Olga_1} and electronic structure calculations by solving a twenty-band \textit{sp$^3$d$^5$s$^*$} tight binding Hamiltonian\cite{Boykin_1}. Both linear and quadratic piezoelectric potentials are calculated by using the published recipe\cite{Usman_4, Schliwa_1} and included in the Hamiltonian. The polarization dependent optical transitions are computed from Fermi's golden rule by absolute values of the optical matrix elements, summed over the spin degenerate states\cite{Usman_1, Usman_5}. The polarization dependent ground state optical transition intensity modes (TE$_{[110]}$, TE$_{[\overline{1}10]}$, and TM$_{[001]}$) are computed as a cumulative sum of the optical transitions between the lowest conduction band state (e1) and the highest five valence band states (h1, h2, h3, h4, and h5)\cite{Usman_1}. 

We want to emphasize here that all of the simulations are performed over very large GaAs buffers surrounding the QDs to properly accommodate the impact of long-range strain and piezoelectric potentials in the electronic structure and the optical transition calculations. For the strain relaxation, we use mixed boundary conditions: bottom fixed, periodic in the lateral directions, and the top free to relax. For the electronic structure calculations, we use closed boundary conditions. The dangling bonds at the surface atoms are passivated according to the published model~\cite{Lee_1}.

\section{Results and Discussions}

In the next two subsections, A and B, we present our results for the flat and tall QDs, respectively, and quantitatively analyse the impact of the elliptical shapes on their electronic and polarization properties. The subsequent subsection C presents results for the vertical stack of nine quantum dot layers (9-VSQDs).

\textit{\textbf{Factors that shifts the electron and hole energies:}} Before we start our analysis of the three QD systems under investigation, we specify the factors that shift the electron and hole energies as the base diameters of the QDs are increased or decreased. We identify four major factors as follows: \\ \\(i) Change in QD Volume ($\bigtriangleup$V): Fig.~\ref{fig:Fig1}(f) provides a quantitative estimate of the changes in the QD volume as a function of the elongation factor $\eta$ for the three types of the elongations. Since the volume only decreases for the Type-I and Type-II elongations, so it will result in an increase of the electron energies and a decrease of the hole energies. \\(ii) Change in QD Diameter ($\bigtriangleup$d): The QD base elongations are based on the increase/decrease of the QD diameters along the [110] and [$\overline{1}$10] directions. While this decrease/increase in the diameters will have very little impact on the ground state electron energy e1 (due to its s-type symmetrical wave function), it will affect the electron and hole p-states energies due to their orientation along these directions. The increase (decrease) in the diameters will produce corresponding increase (decrease) in the electron energies and decrease (increase) in the hole energies, oriented along their directions. \\(iii) Strain: The strain directly modifies the band edges and thus impacts the electron and hole confinement energies. The electron energies are shifted by changes in the hydrostatic strain ($\epsilon_{xx}+\epsilon_{yy}+\epsilon_{zz}$) only, whereas the hole energies are affected by changes in both the hydrostatic and the biaxial strain ($\epsilon_{xx}+\epsilon_{yy}-2\epsilon_{zz}$). Simple analytical relations based on the deformation potential theory can be applied to estimate these changes\cite{Usman_3}. \\(iv) Piezoelectric Potential: InAs/GaAs systems are strongly piezoelectric: the orientation and the magnitude of the piezoelectric potentials significantly impacts the orientation and the splitting of the electron and hole excited states. It should be noted that although the piezoelectric potentials do not directly shift the electron and hole p-state energies, but they determine the $\bigtriangleup$d induced changes by controlling the orientation of the p-states.

Overall for the Type-I elongation, $\bigtriangleup$V is large whereas $\bigtriangleup$d is small since we keep one diameter unchanged and only reduce the other diameter by $\bigtriangleup$d, so the impact of $\bigtriangleup$V dominates. For the Type-II elongation, the QD volume only slightly decreases, so the impact of $\bigtriangleup$V is small. However, we increase one diameter by $\bigtriangleup$d and decrease the other diameter by $\bigtriangleup$d, so the overall impact of $\bigtriangleup$d is much stronger. Finally, for the Type-IIv elongation, $\bigtriangleup$V=0, and $\bigtriangleup$d is roughly same as for the Type-II elongation.                    

\subsection{Flat Quantum Dot (AR=0.225)}

In this subsection, we study the impact of the elongations on the electronic and polarization properties of a flat QD as shown by the schematic in Fig.~\ref{fig:Fig1}(a). The QD has a base diameter of 20 nm and a height of 4.5 nm (AR = 5/20 = 0.225). Such low AR QDs are more commonly obtained from the strain-driven SK self-assembly growth process and their electronic properties have been widely studied in the literature.

\begin{figure*}
\includegraphics[scale=0.28]{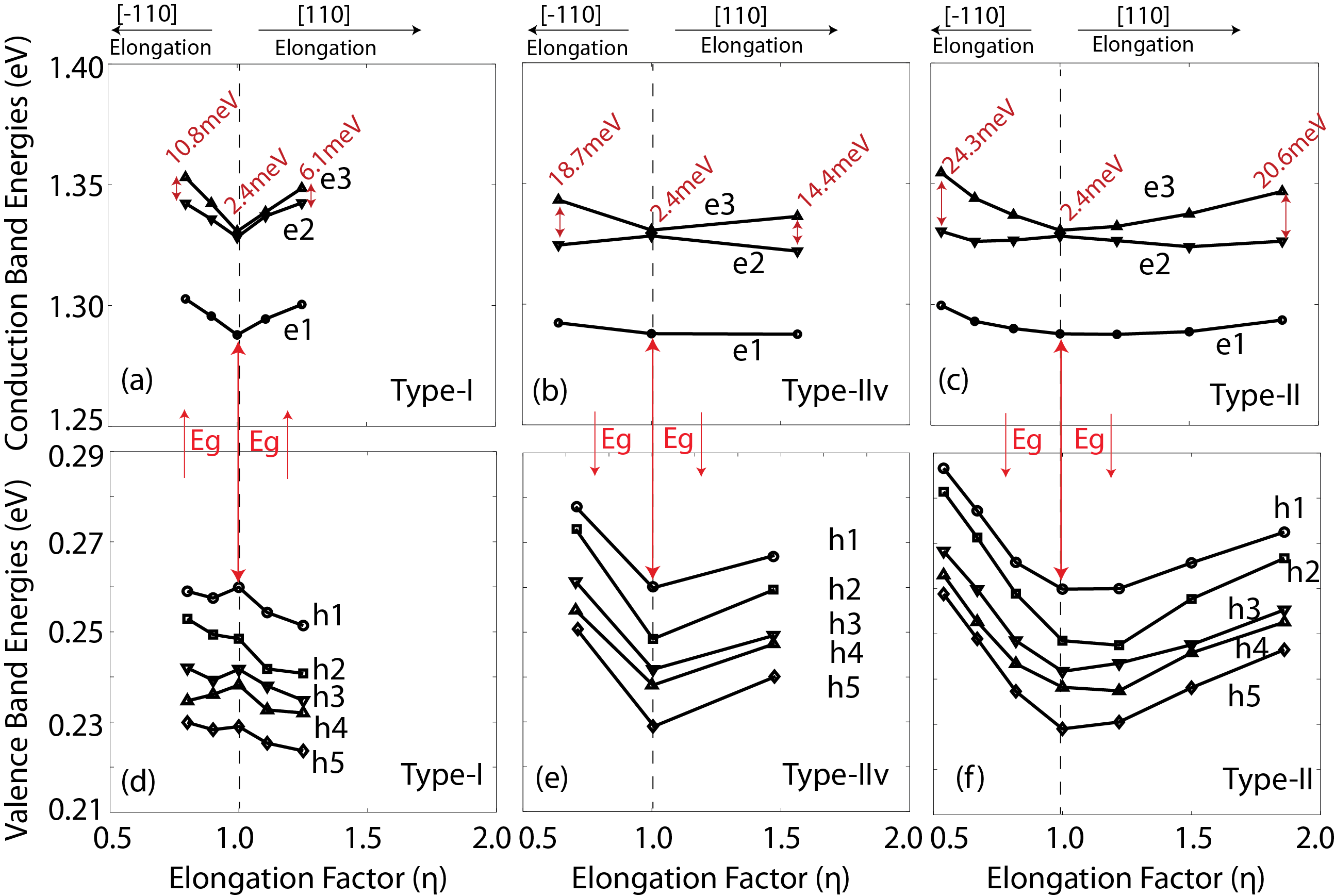}
\caption{(a, b, c) The lowest three conduction band energy levels (e1, e2, and e3) are plotted as a function of the elongation factor ($\eta$) for the (a) Type-I, (b) Type-IIv, and (c) Type-II elongations. (c, d, e) The highest five valence band energy levels (h1, h2, h3, h4, and h5) are plotted as a function of the QD elongation factor ($\eta$) for the (a) Type-I, (b) Type-IIv, and (c) Type-II elongations. The corresponding increase/decrease in the optical gap energy (E$_{g}$) is also specified in each case by using the vertical arrows.}
\label{fig:Fig2}
\end{figure*}
\vspace{1mm}  

\begin{figure*}
\includegraphics[scale=0.16]{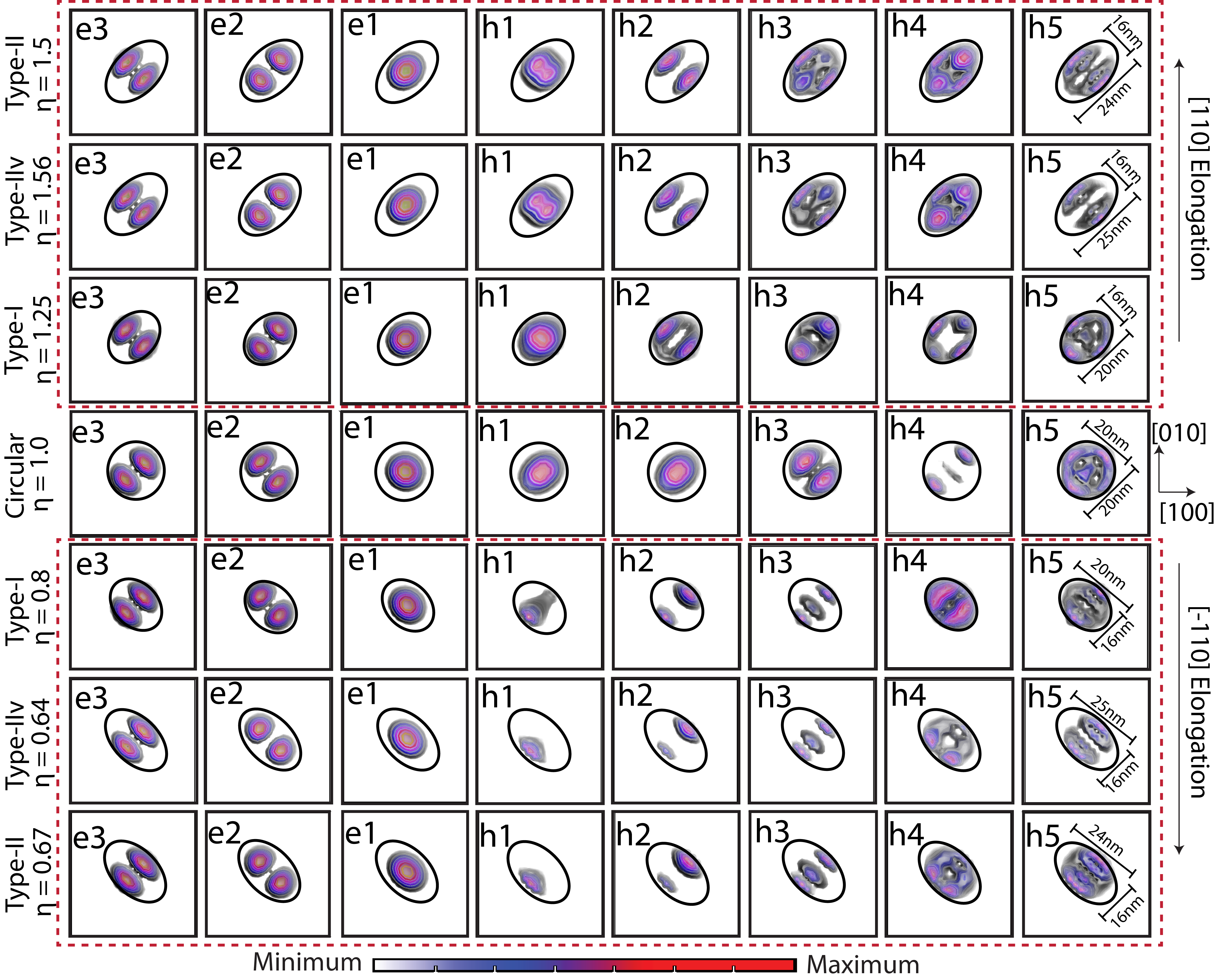}
\caption{The top view of the wave function plots for the lowest three conduction band (e1, e2, and e3) and the highest five valence band (h1, h2, h3, h4, and h5) states are shown for the circular-base QD and for the selected elongations of the QD. The intensity of the colors in the plots represent the magnitude of the wave functions, with the dark red color indicating the largest magnitude and the light blue color indicating the smallest magnitude. The boundaries of the QDs are also shown to guide the eye.}
\label{fig:Fig3}
\end{figure*}
\vspace{1mm} 

\begin{SCfigure*}
\includegraphics[scale=0.3]{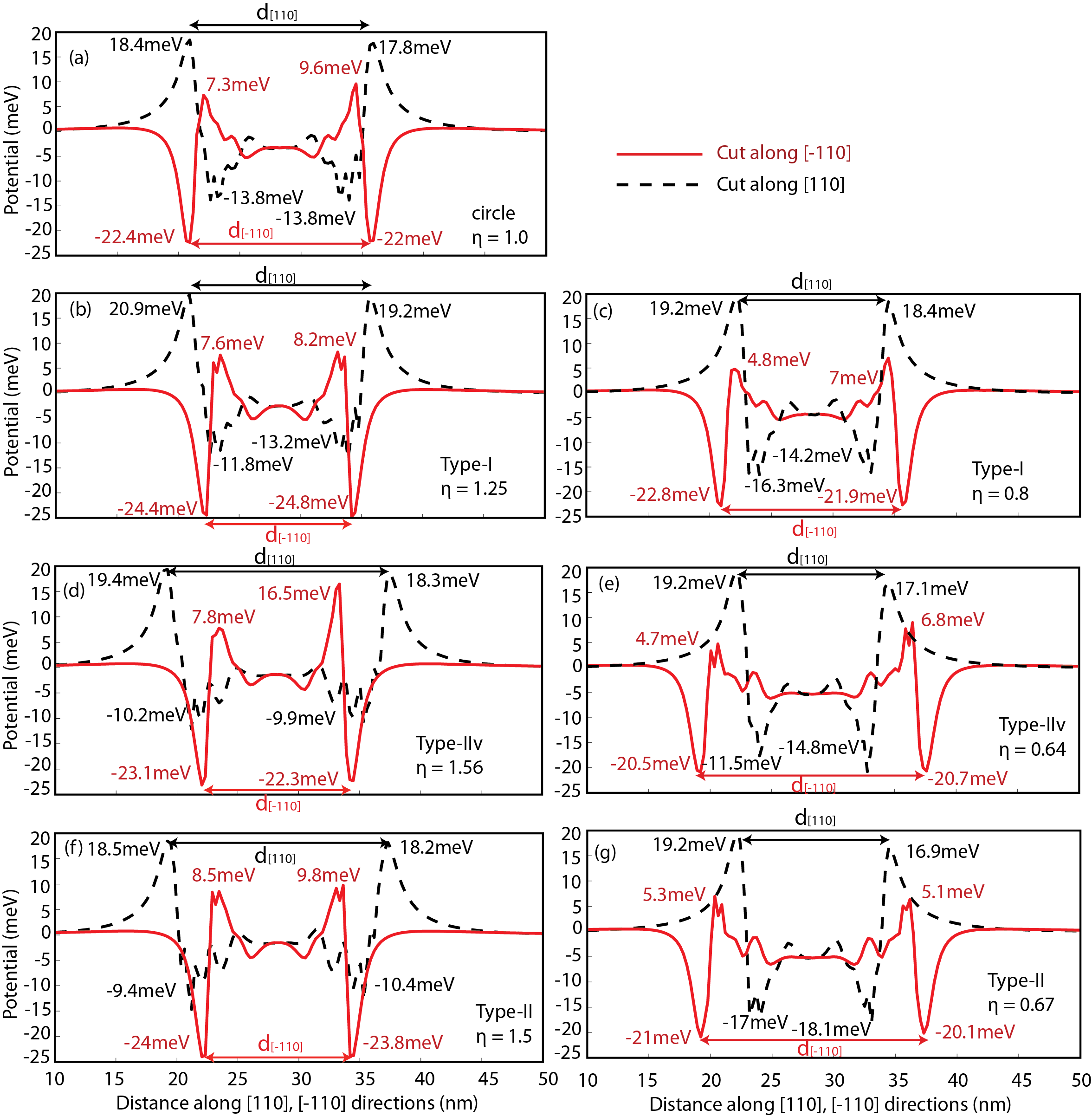}
\caption{The plots of the total piezoelectric potentials (linear+quadratic) are shown for (a) circular-base QD, (b, d, f) [110]-elongated QDs, and (c, e, g) [$\overline{1}$10]-elongated QDs. In each case, the type and magnitude of elongation is specified. The solid red lines are plotted along the [$\overline{1}$10] direction through the center of the QD, 0.5 nm above its base. The dotted (broken) black lines are plotted along the [110] direction through the center of the QD, 0.5 nm above its base. The boundaries of the QD region are also marked in each case by specifying the lengths of QD along the [110] and [$\overline{1}$10] directions, d$_{[110]}$ and d$_{[\overline{1}10]}$.}
\label{fig:Fig4}
\end{SCfigure*}
\vspace{1mm} 

\subsubsection{Electronic properties of the flat QD}

Fig.~\ref{fig:Fig2} plots the lowest three conduction band energy levels (e1, e2, e3) and the highest five valence band energy levels (h1, h2, h3, h4, h5) for the three types of elongations: (a, d) Type-I, (b, e) Type-IIv, and (c, f) Type-II. The figures are plotted using the same scale to facilitate mutual comparison. In order to understand the shifts in the energies, we also plot wave functions and piezoelectric potentials for the few selected cases. Fig.~\ref{fig:Fig3} shows the top view of the wave function plots for the circular-base QD and for the few selected [110] and [$\overline{1}$10] elongations. The QD boundaries and dimensions are also marked in each case. 

Fig.~\ref{fig:Fig4} plots the total (linear+quadratic) piezoelectric potentials along the [110] (dotted lines) and [$\overline{1}$10] (solid lines) directions through the center of the QDs, $\approx$0.5 nm above their base. The quadruple nature of the potentials is clearly evident which has been well established in the literature\cite{Usman_4, Schliwa_1, Islam_1} and is shown to strongly influence the orientation of the electron and hole p-states. In this case of the flat QD with AR=0.225, we find that the quadratic component of the piezoelectric potential does not fully cancel the linear component inside the QD region in contrast to the previous \textbf{k}$\centerdot$\textbf{p} study~\cite{Schliwa_1}, where the quadratic and the linear components were found to fully cancel each other for the AR $<$ 0.5. Since the piezoelectric potentials are a strong function of the QD shape and composition, so we find that their results cannot be generalized to all types of QDs. Theoretical studies by Usman \textit{et al}.\cite{Usman_4} and Islam \textit{et al}.\cite{Islam_1} have also shown non-zero values of the net piezoelectric potentials for the similar QDs. 

Our calculations show that the piezoelectric potential plots have two peaks at the QD interfaces, one just outside the QD region and one just inside the QD region. The electron and hole wave functions are found to be more influenced by the piezoelectric potential peaks inside the QD, which causes the lower electron p-state (e2) to align along the [$\overline{1}$10] direction and the ground hole state (h1) to slightly elongate along the [110] direction for the circular-base QD (see Fig.~\ref{fig:Fig3} for $\eta$ = 1.0). The alignment of the lower electron p-state (e2) along the [$\overline{1}$10] direction is in agreement with the experimental reports for the similar QDs.\cite{Boucaud_1, Maltezopoulos_1}    

\textbf{\textit{Lowest three conduction band energies:}} From our atomistic relaxations, we find that the hydrostatic component ($\epsilon_{xx}+\epsilon_{yy}+\epsilon_{zz}$) of the strain remains unchanged for both the [110] and the [$\overline{1}$10] elongations. Since the electron energies are only affected by the hydrostatic strain, so the strain does not contribute in the shifts of the conduction band energies. Also shown in Fig.~\ref{fig:Fig4} that the piezoelectric potentials do not exhibit any significant change for the elongated QDs. The peaks outside the QD regions are only changed by 1 to 2 meV and the peaks inside the QD region are decreased by 2 to 4 meV. These relatively small changes in the potentials will not result in any noticeable effects on the e2 and e3 energies. Therefore, we conclude that the electron energies are mainly affected by the changes in the diameters ($\bigtriangleup$d) and the volume ($\bigtriangleup$V) of the flat QD, while the strain and the piezoelectric fields have only minor contributions as a function of $\eta$.

The lowest electron energy level (e1) has s-type symmetrical wave function and is mainly affected by the decrease in the QD volume ($\bigtriangleup$V). The QD diameter change ($\bigtriangleup$d) does not have any noticeable impact on its energy. In Fig.~\ref{fig:Fig2}(a), as the QD volume decreases for the [110] and the [$\overline{1}$10] elongations, a nearly symmetric increase in the e1 energy is calculated. When the volume of the QD is kept fixed as in Fig.~\ref{fig:Fig2}(b), e1 energy is almost unchanged confirming that $\bigtriangleup$d has only minor impact on e1. Finally, for the Type-II elongation (Fig.~\ref{fig:Fig2}(c)), $\bigtriangleup$V is very small and a corresponding small increase in the e1 energy is observed. Fig.~\ref{fig:Fig3} shows that the wave function for the e1 state also retains its s-type symmetry, with only slight elongation along the direction in which the QD is being elongated.       

The excited conduction band states (e2 and e3) have p-type symmetry and thus gets strongly affected by all of the three types of elongations. The $\bigtriangleup$V is once again a dominant factor, which pushes these energy levels towards higher values. This is evident from Fig.~\ref{fig:Fig2}(a) where both e2 and e3 increase in energy irrespective of the elongation direction. However, if $\bigtriangleup$V=0 as in Fig.~\ref{fig:Fig2}(b), the increase in the QD diameter reduces the energy of the state aligned along its direction and vice versa.

The type-II elongation is an interesting case where the two factors, $\bigtriangleup$V and $\bigtriangleup$d, compete as the QD diameters along the [110] and [$\overline{1}$10] directions are changed by equal values. Since the lower p-state (e2) is always oriented along the major-axis of the elliptical shape, so any increase in the corresponding diameter tends to reduce its energy while the decrease in the QD volume pushes it towards the higher energies. For the small values of the elongation factor ($\eta$ = 0.67, 0.82, 1.22, and 1.5), e2 decreases due to dominance of $\bigtriangleup$d induced shift. However, for the larger values of the  elongation factor ($\eta$ = 0.54 and 1.86), the $\bigtriangleup$V induced upward shift overcomes the $\bigtriangleup$d induced downward shift and hence the energy of e2 increases. 

The energy of the higher p-state, e3, always increases as a function of $\eta$ because it is oriented along the shorter diameter direction and hence an increase in its energy is supported by both, $\bigtriangleup$V and $\bigtriangleup$d.

\textbf{\textit{Electron p-state splitting:}} The energy difference between the p-states ($\bigtriangleup$e$_p$ = e3-e2) is an important parameter of interest as it provides a measure of the confinement anisotropy between the [110] and [$\overline{1}$10] directions\cite{Schliwa_1} and is sometimes used to characterize the fine structure splitting (FSS)\cite{Singh_2}. We find that $\bigtriangleup$e$_p$ is always larger for the [$\overline{1}$10]-elongations as compared to the [110]-elongations for the same of $\eta$. This is because for the circular-base QD ($\eta$ = 1.0), the cumulative effect of the underline zincblende crystal asymmetry, strain, and piezoelectricity results in $\bigtriangleup$e$_{p} \approx$ 2.4 meV and favours [$\overline{1}$10] direction for the e2 state (see Fig.~\ref{fig:Fig3} for $\eta$ = 1.0). Therefore, any [$\overline{1}$10] elongation merely enhances this asymmetry, whereas a [110] elongation first needs to overcome this inherent $\approx$2.4 meV splitting in order to flip the orientation of e2 state and hence results in overall lower values of $\bigtriangleup$e$_p$.

\textbf{\textit{Separation between the lowest two electron energies:}} Another parameter of interest for the laser design is the difference between the lowest two conduction band energy levels ($\bigtriangleup$e$_{21}$ = e2-e1) which should be large to avoid undesirable occupancy of the excited states\cite{Usman_3}. The largest reductions in $\bigtriangleup$e$_{21}$ are calculated to be $\approx$1 meV, $\approx$8 meV, and $\approx$10 meV for the Type-I, Type-IIv, and Type-II elongations, respectively. These small variations in $\bigtriangleup$e$_{21}$ suggest that the elongations of a flat QD does not deteriorate this parameter for the implementation of laser operation.          

\textbf{\textit{Highest five valence band energies:}} Figs.~\ref{fig:Fig2}(d), (e), and (f) plot the highest five valence band energy levels (h1, h2, h3, h4, h5) for the three types of the elongations. The shifts in the valence band energy levels are more complicated to understand due to their intermixed HH/LH characters and much stronger confinements within the QD region. The changes in their energies exhibit drastic differences for the three types of the elongations.

Due to the heavier effective mass and stronger confinement inside the QD region, the orientation of the hole wave functions are mainly determined by the piezoelectric potential peaks inside the QD region. In the case of the circular-base QD as shown in Fig.~\ref{fig:Fig3}, h1 and h2 are slightly elongated towards the [110]-direction due to the negative peaks of the piezoelectric potentials along this direction inside the QD region.

For the [110]-elongations, the piezoelectric potential inside the QD region slightly reduces and the decrease in d$_{[\overline{1}10]}$ diameter results in an enhanced impact of the larger negative peaks of the potential outside the QD along the [$\overline{1}$10] direction. These two factors favour forces the hole wave functions to align along the [$\overline{1}$10] direction as shown in Fig.~\ref{fig:Fig3}. When the QD is elliptical with its major-axis along the [$\overline{1}$10] direction, the internal negative piezoelectric potential slightly increases and dominate to align the hole wave functions along the [110] direction. Overall, we find that the hole wave functions always tend to align along the minor-axis of the elliptical flat QD.

\begin{figure*}
\includegraphics[scale=0.4]{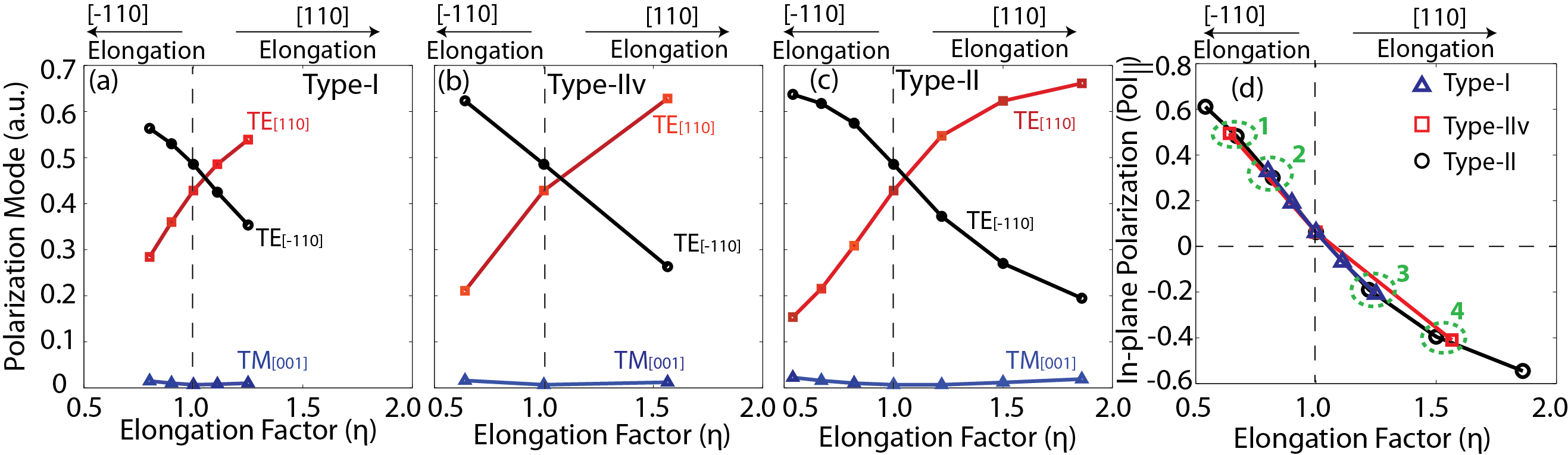}
\caption{The polarization dependent optical transition modes TE$_{[110]}$, TE$_{[\overline{1}10]}$, and TM$_{[001]}$ are plotted as a function of the (a) Type-I, (b) Type-IIv, and (c) Type-II QD elongations. The figures are plotted using the same scales to facilitate easy mutual comparison. (d) The plots of the in-plane polarization anisotropy (Pol$_{||}$) as defined by Eq.~\ref{eq:pol} are shown for the three types of elongations, exhibiting an inverse quadratic dependence on $\eta$, consistent with the Sheng \textit{et al.}\cite{Sheng_1}. Four cases are marked by using green ovals indicating that the two different types of elongations, with roughly similar values of $\eta$, exhibiting similar values of the Pol$_{||}$.}
\label{fig:Fig5}
\end{figure*}
\vspace{1mm}   

The biaxial strain for the Type-I elongation slightly relaxes and along with large $\bigtriangleup$V, it dominates the upward shift induced by $\bigtriangleup$d and pushes the hole energies towards lower values. This is clearly evident for the [110] elongations. However for the [$\overline{1}$10] elongations, since the hole wave functions exhibit stronger alignment along the [110] direction , the increase in energies coming from $\bigtriangleup$d is enhanced and therefore some of the hole energies can be observed shifting slightly upward in the Fig.~\ref{fig:Fig2}(d).  

For the Type-IIv elongation in Fig.~\ref{fig:Fig2}(e), $\bigtriangleup$V = 0 and thus the impact of $\bigtriangleup$d, which is much larger than for the Type-I elongation, dominates. We also find that the biaxial strain relaxation is much weaker in this case, and thus its downward shift is also very small. As a result, the hole energies shift towards larger values. 

Finally, $\bigtriangleup$V is very small for the Type-II elongation, and $\bigtriangleup$d is very large, so overall its impact dominates and shifts all the hole energies towards higher values. The impact of the biaxial strain relaxation is again very small downward shift. 

As a summary, for the flat QD under elongations, the shifts in the electron and hole energies are mainly governed by $\bigtriangleup$V and $\bigtriangleup$d, whereas the direct impact of the strain remains negligibly small.     

\textbf{\textit{Optical gap energy, E$_{g}$:}} The optical gap energy (E$_g$ = e1-h1) increases for both, the [110] and the [$\overline{1}$10], Type-I elongations, thus blue shifting the ground state optical wavelength mainly due to a large decrease in the QD volume. However, if the QD volume is fixed or only slightly decreased as for the Type-IIv and Type-II elongations, respectively, E$_{g}$ decreases as a function of the elongation and hence results in a red shift of the ground state optical transition wavelength.

\subsubsection{Polarization properties of the flat QD}

Figs.~\ref{fig:Fig5}(a), (b), and (c) compare polarization dependent TE and TM modes for the three QD elongations under study. The elliptical shape of the QD, irrespective of the orientation of its major-axis, tends to increase the TE-mode along its major-axis and decreases the TE-mode along its minor-axis. This can be understood as follows: the few top most valence band states have dominant heavy hole (HH) character due to the strain induced large splitting between the HH and LH bands. These heavy hole states are mainly comprised of $|X\rangle$ and $|Y\rangle$ symmetry wave functions, where $X$ and $Y$ are selected along the high symmetry [110] and [$\overline{1}$10] directions, respectively. The lowest electron state (e1) is mainly symmetric $|S\rangle$ type wave function (as a good approximation). The elongation of QD along, for example, the $X$-direction will have negligible impact on the $|S\rangle$ type wave function, but it will increase (decrease) $|X \rangle$ ($|Y\rangle$) component of the valence band states. Therefore, the TE$_{X} \propto |\langle X|S\rangle|^2$ component of the electron-holes transition will increase and the TE$_{Y} \propto |\langle Y|S\rangle|^2$ component will decrease as evident from Figs.~\ref{fig:Fig5}(a), (b), and (c). 

\begin{table*}
\caption{\label{tab:table1} The ratio of the LH components in the highest five valence band states (h1, h2, h3, h4, h5) of the elliptical-shaped flat QD with respect to the LH components of the corresponding valence band states of the circular-base ($\eta$) QD for a few selected values of the elongation factor, $\eta$. For each case, the corresponding ratio of the electron-hole wave function spatial overlap along the $\vec{z}$ = [001] direction given by e1hi = $| \langle \psi_{e1} | \vec{z} | \psi_{hi} \rangle |$ is also provided.}
\renewcommand{\arraystretch}{1.5}
\begin{tabular}{|l|l|l|l|l|l|l|}
\multicolumn{7}{c}{} \\
\cline{1-7}
\multicolumn{1}{|c|}{\textbf{Elongation type}} & 
\multicolumn{1}{c|}{\textbf{$\pmb{\eta}$}} & 
\multicolumn{1}{c|}{\textbf{h1 (e1h1)}} &
\multicolumn{1}{c|}{\textbf{h2 (e1h2)}} &
\multicolumn{1}{c|}{\textbf{h3 (e1h3)}} &
\multicolumn{1}{c|}{\textbf{h4 (e1h4)}} &
\multicolumn{1}{c|}{\textbf{h5 (e1h5)}} \\ 
\cline{1-7}
\multicolumn{1}{|c|}{\multirow{2}{*}{\textbf{Type-I}}} & 1.25 & 1.38 (4.12) & 1.49 (12.88) & 1.15 (1.49) & 0.97 (1.07) & 0.90 (0.57) \\ 
\cline{2-7}
\multicolumn{1}{|c|}{} & 0.80 & 1.62 (5.05) & 1.19 (13.59) & 1.27 (0.99) & 1.11 (11.53) & 1.04 (2.49) \\ 
\cline{1-7}
\multicolumn{1}{|c|}{\multirow{2}{*}{\textbf{Type-IIv}}} & 1.565 & 1.31 (7.59) & 1.29 (2.76) & 1.33 (0.22) & 0.96 (1.31) & 1.14 (0.34) \\ 
\cline{2-7}
\multicolumn{1}{|c|}{} & 0.64 & 1.56 (7.71) & 1.05 (17.76) & 1.09 (1.48) & 0.97 (3.96) & 0.90 (1.96) \\ 
\cline{1-7}
\multicolumn{1}{|c|}{\multirow{2}{*}{\textbf{Type-II}}} & 1.86 & 1.68 (9.59) & 1.39 (4.94) & 1.45 (0.77) & 1.00 (0.93) & 1.20 (1.96) \\ 
\cline{2-7}
\multicolumn{1}{|c|}{} & 0.54 & 1.81 (9.71) & 1.12 (23.76) & 1.15 (2.00) & 1.05 (2.38) & 0.94 (0.81) \\ 
\cline{1-7}
\end{tabular} 
\end{table*}

The analysis of the calculated TM$_{[001]}$ component reveals that it also increases for the elliptical QDs. The dominant contribution in the ground state optical intensity for the flat-shaped QDs comes from the highest valence band state (h1), with the lower valence band states (h2, h3, h4, and h5) only adding weak transition strengths\cite{Usman_5}. The strength of the TM$_{[001]}$ component is directly related to the LH mixing in the valence band states and its magnitude is proportional to the electron-hole wave function spatial overlap along the growth ([001]) direction, given by e1hi = $| \langle \psi_{e1} | \overrightarrow{z} | \psi_{hi} \rangle |$. Here $\psi_{e1}$ is the ground electron state, $\psi_{hi}$ is the $i$th valence band state where $i \in \{$ 1, 2, 3, 4, 5$\}$, and $\overrightarrow{z}$ is along the [001]-direction.

Table~\ref{tab:table1} provides values of the ratios of the LH component in the top five valence band states for the few selected elliptical shapes with respect to the circular-base shape. For each case, the corresponding ratio of the spatial overlap (e1hi) is also provided with in the (). Our calculations show an increasing LH mixing in the valence band states for the elliptical QDs with respect to the circular-base QD, in particular for the h1 state which gives dominant e1-h1 transition. The spatial overlap between electron and hole wave funcitons also increases and therefore, a net increase in the TM$_{[001]}$ mode is calculated as a function of $\eta$. This characteristic of the pure InAs QDs is in contrast to the In$_{0.5}$Ga$_{0.5}$As ordered and disordered QDs reported by Singh \textit{et al.}\cite{Singh_1}, where they have shown that the LH character of the h1 valence band state remains unchanged as a function of the QD elongation. Based on these results, we predict that the elliptical shape has a stronger impact on the polarization response of the pure InAs QDs, as compared to the alloyed InGaAs QDs.     

\begin{figure*}
\includegraphics[scale=0.3]{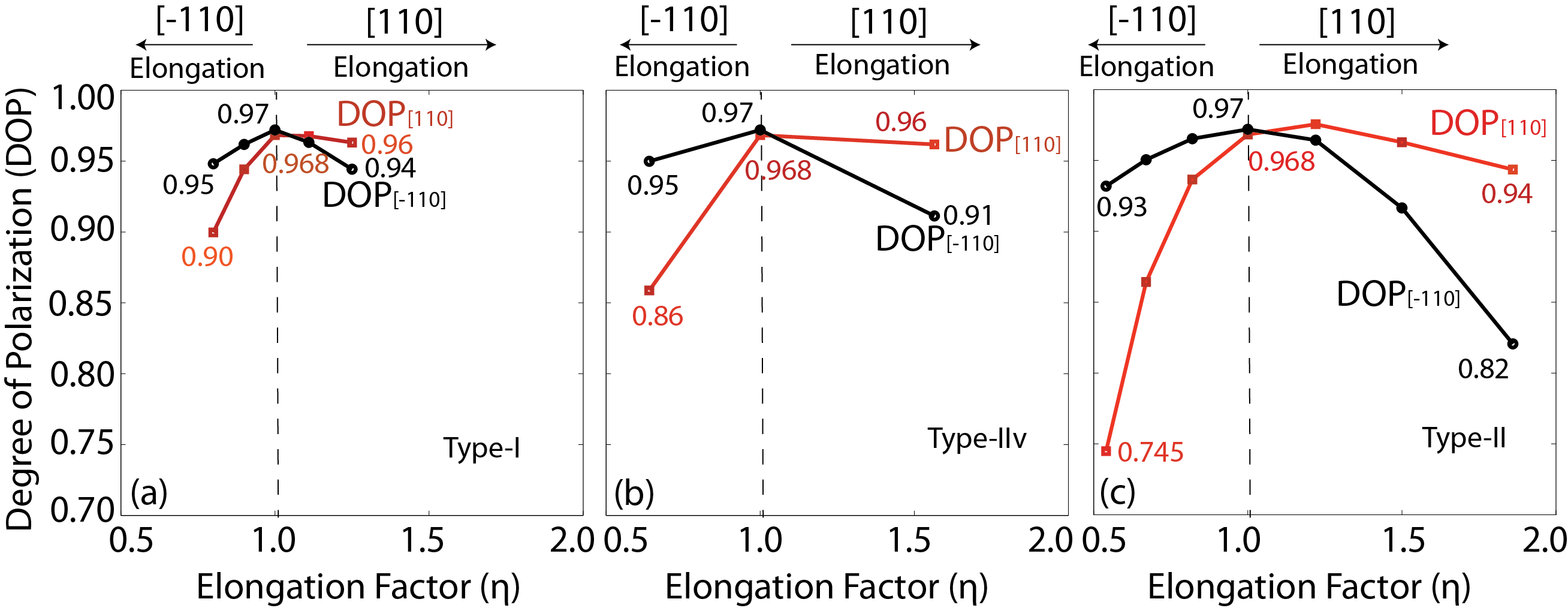}
\caption{The plots of the DOP$_{[110]}$ and DOP$_{[\overline{1}10]}$ are shown as a function of the (a) Type-I, (b) Type-IIv, and (c) Type-II QD elongations. The values of the DOP decrease irrespective of the direction of the QD elongation.}
\label{fig:Fig6}
\end{figure*}
\vspace{1mm}    

\textbf{\textit{In-plane polarization anisotropy:}} Fig.~\ref{fig:Fig5}(d) plots the in-plane polarization anisotropy (Pol$_{||}$) as defined by Eq.~\ref{eq:pol} for the three types of the elongations. We find that our results for Pol$_{||}$ are overall in agreement with the findings of Sheng \textit{et al.}\cite{Sheng_1}, that when the height of the QD is kept fixed, similar elongations ($\eta$) results in nearly similar values of the in-plane anisotropy. We highlight four such cases in Fig.~\ref{fig:Fig5}(d) using green dotted circles where nearly same values of Pol$_{||}$ are calculated for the following sets of values of $\eta$: (1) 0.64 in Type-IIv and 0.67 in Type-II, (2) 0.82 in Type-II and 0.8 in Type-I, (3) 1.22 in Type-II and 1.25 in Type-I, and (4) 1.565 in Type-IIv and 1.50 in Type-II. Therefore, we conclude that the in-plane polarization anisotropy (Pol$_{||}$) only depends on the value of $\eta$, irrespective of the type of the elongation provided that the height of the QD is kept fixed. We also find that our calculated values of Pol$_{||}$ for all of the three type of the elongations roughly follow an inverse quadratic dependence on the elongation factor ($\eta$), in agreement with the quadratic dependence on the lateral aspect ratio ($\beta$) reported by Sheng \textit{et al.}\cite{Sheng_1}, as by definition $\eta$ = $\beta^{-1}$. 

\textbf{\textit{Tuning of DOP$_{[\overrightarrow{n}]}$:}} Figs.~\ref{fig:Fig6}(a), (b), and (c) investigate the changes in the DOP$_{[\overrightarrow{n}]}$ for the three types of the elliptical shapes. Significant changes in the values of the DOP$_{[\overrightarrow{n}]}$ are observed as a function of $\eta$. More interestingly, it is found that both DOP$_{[110]}$ and DOP$_{[\overline{1}10]}$ decrease irrespective of the type and direction of the elongation. This implies that the elliptical-shape, in general, improves the polarization response compared to the circular-base for the flat QD. The largest decrease ($\approx$23\%) in the value of the DOP$_{[110]}$ (from 97$\%$ to 74.5$\%$) is calculated for the [$\overline{1}$10] elongation ($\eta$ = 0.54).
       
\subsection{Tall Quantum Dot (AR=0.40)}

In this subsection, we study the impact of elliptical shapes on the electronic and polarization properties of a tall InAs QD, having the same base diameter (20 nm) as of the flat QD of the previous subsection, but with a height of 8 nm (AR = h/d = 8/20 = 0.40). Such high AR QDs are obtained using special growth conditions (very slow growth rate and high temperatures)\cite{Bimberg_1}, or are typically found in the optically active upper layers of the weakly coupled bilayer QD stacks\cite{Usman_2}, where the presence of strain from the lower QD layers results in the larger size of the upper layer QDs. To our knowledge, no detailed theoretical investigation of the polarization properties of such tall elliptical dome-shaped QDs is available in the literature, as the previous theoretical studies have only focused on the flat shapes of the QDs with low ARs: AR = 2/20 = 0.1~\cite{Favero_1}, AR $\approx$ 0.17~\cite{Pryor_1}, AR = 3.5/20 = 0.175~\cite{Singh_2}, AR = 2/25 = 0.08~\cite{Mlinar_1}, and AR = 4.5/28.8 = 0.16~\cite{Sheng_1}. 

Schliwa \textit{et al}.~\cite{Schliwa_1} applied \textbf{k}$\centerdot$\textbf{p} theory to study the electronic and optical properties of the QDs by varying their AR from 0.17 to 0.5. However in order to vary the AR, they choose to keep the QD volume constant by simultaneously changing both the base diameter and the height of the QD. Therefore their results do not isolate the impact of only varying the lateral aspect ratio. Their study of the base elongations which is focused at the pyramidal-shaped QDs shows that the linear and quadratic piezoelectric potentials fully cancel each other for all values of the AR $<$ 0.5, whereas our atomistic simulations presented in the previous subsection have already shown a non-zero net piezoelectric potential in the interior of a dome-shaped QD with the AR = 0.225. We therefore conclude that their result can not be generalized for all types of the QDs. In order to extend our study of the previous subsection, here we investigate a tall QD by just increasing the height of the QD from 4.5 nm to 8 nm, keeping its base diameter same (20 nm) as for the flat QD. This allows us to make a direct comparison between the results for the flat and the tall QDs. Once again, for all types of the elongations, the height of the QD is kept constant at 8 nm.

\subsubsection{Electronic properties of the tall QD}

\begin{figure*}
\includegraphics[scale=0.28]{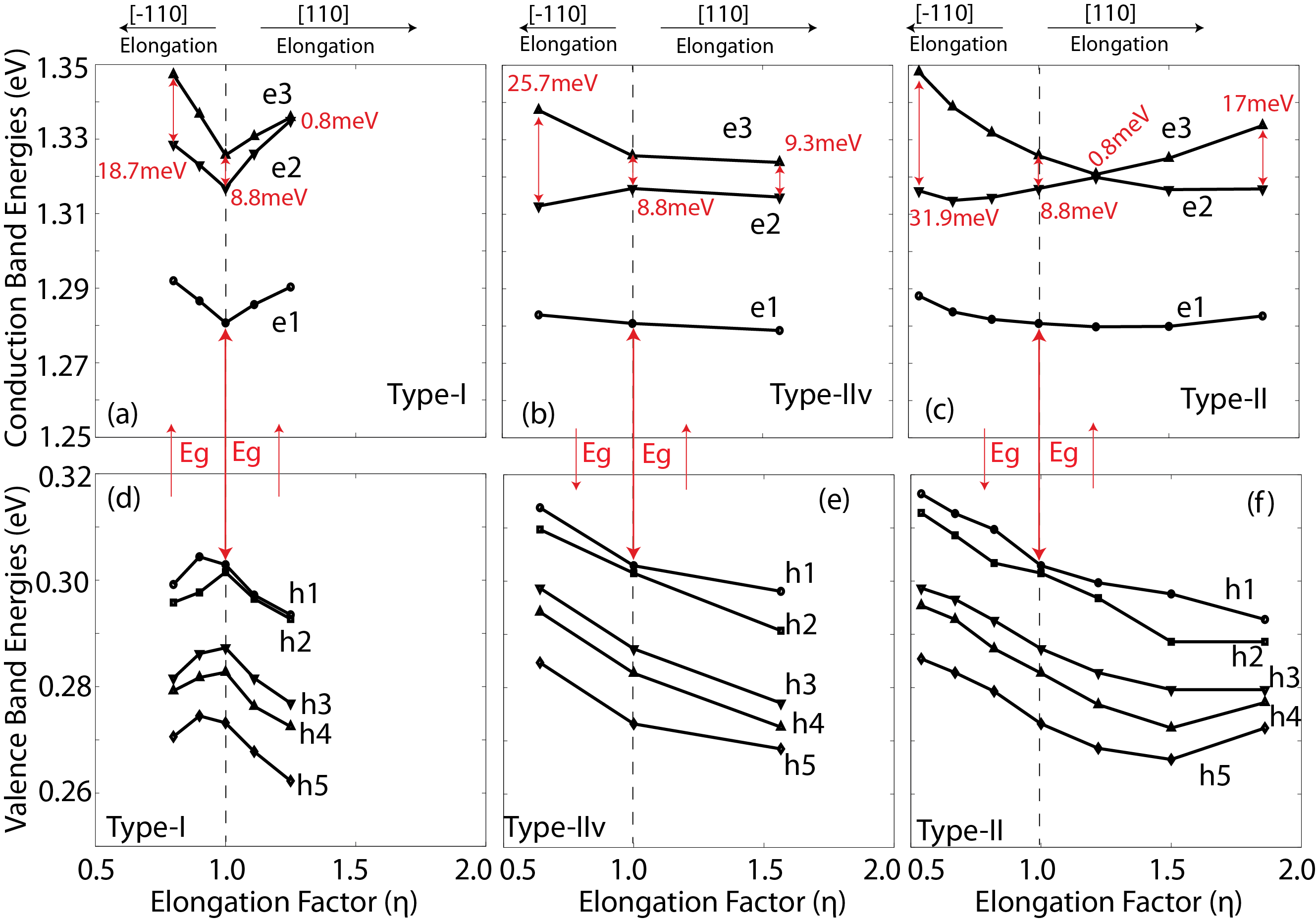}
\caption{(a, b, c) The lowest three conduction band energy levels (e1, e2, and e3) are plotted as a function of the elongation factor ($\eta$) for the (a) Type-I, (b) Type-IIv, and (c) Type-II elongations. (c, d, e) The highest five valence band energy levels (h1, h2, h3, h4, and h5) are plotted as a function of the QD elongation factor ($\eta$) for the (a) Type-I, (b) Type-IIv, and (c) Type-II elongations. The corresponding increase/decrease in the optical gap energy (E$_{g}$) is also specified in each case by using the vertical arrows.}
\label{fig:Fig7}
\end{figure*}

\begin{figure*}
\includegraphics[scale=0.15]{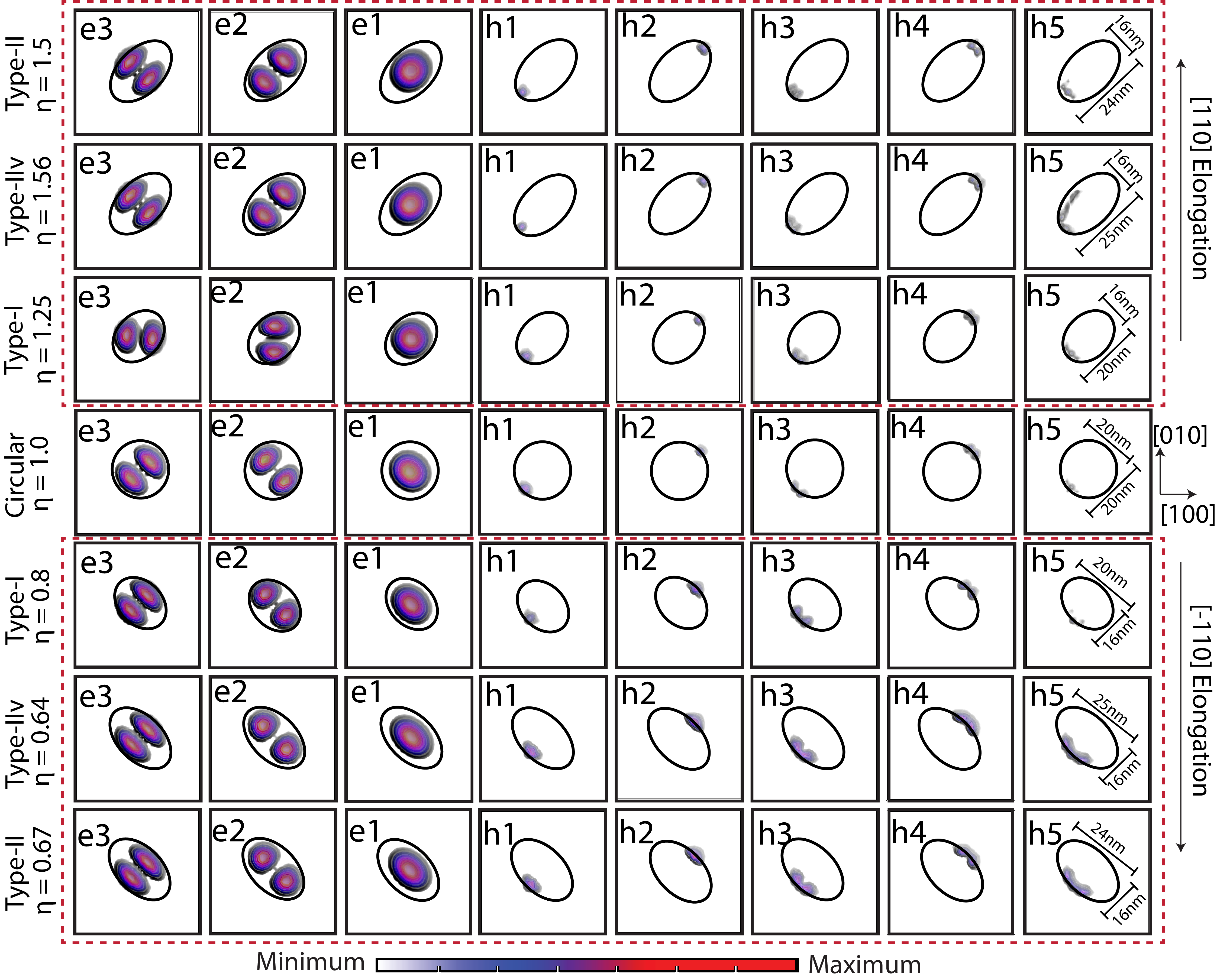}
\caption{The top view of the wave function plots for the lowest three conduction band (e1, e2, and e3) and the highest five valence band (h1, h2, h3, h4, and h5) states are shown for the circular-base of the QD and for the selected elongations of the QD. The intensity of the colors in the plots represent the magnitude of the wave functions, with the dark red color indicating the largest magnitude and the light blue color indicating the smallest magnitude. The boundaries of the QDs are also shown to guide the eye.}
\label{fig:Fig8}
\end{figure*}

\begin{SCfigure*}
\includegraphics[scale=0.3]{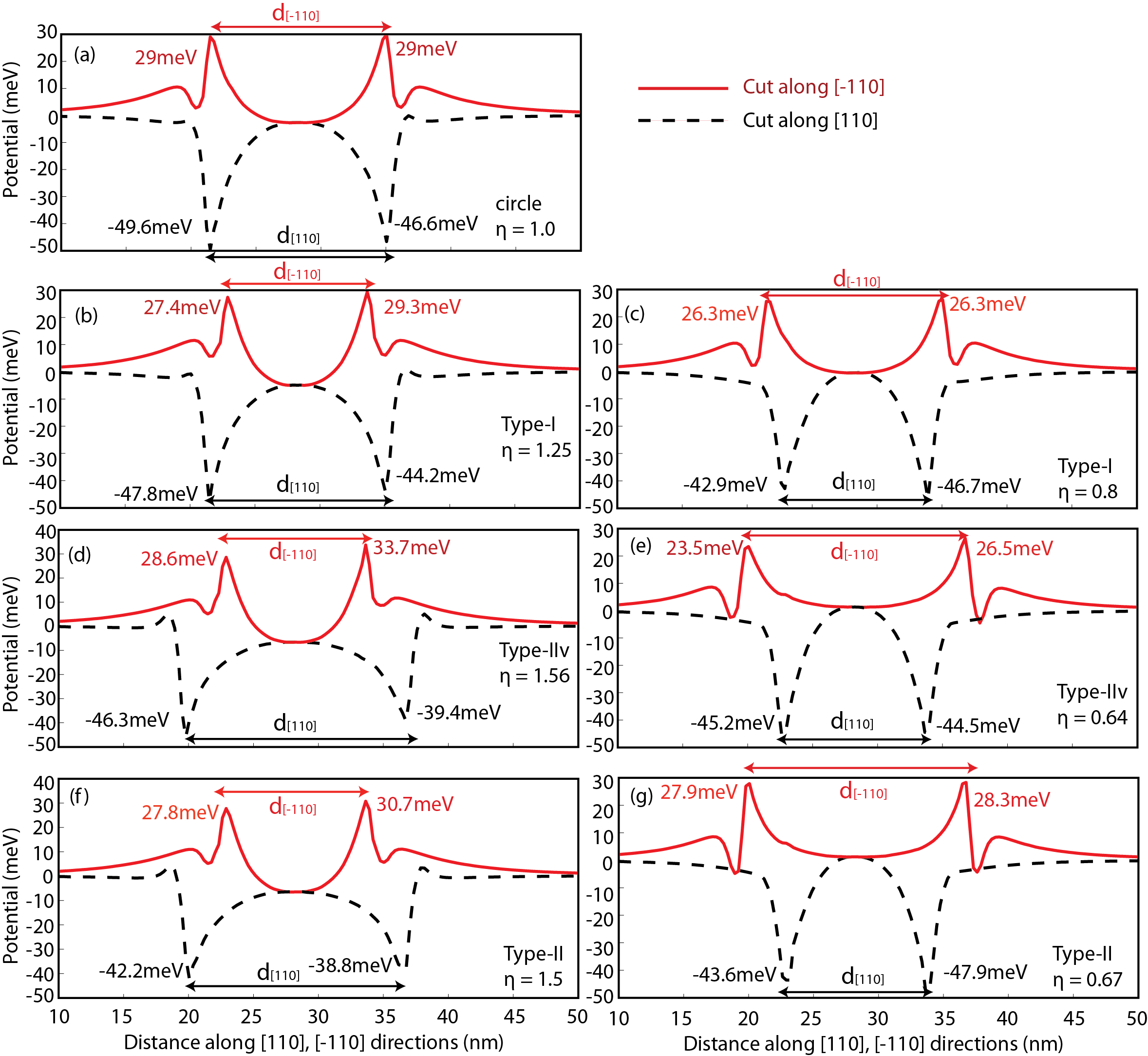}
\caption{The plots of the total (linear+quadratic) piezoelectric potentials are shown for (a) circular-base QD, (b, d, f) [110]-elongated QDs, and (c, e, g) [$\overline{1}$10]-elongated QDs. In each case, the type and the magnitude of the elongation is specified. The solid red lines are plotted along the [$\overline{1}$10] direction through the center of the QD, 0.5 nm above its base. The dotted (broken) black lines are plotted along the [110] direction through the center of the QD, 0.5nm above its base. The boundary of the QD region is also marked in each case by specifying the lengths of QD along the [110] and [$\overline{1}$10] directions, d$_{[110]}$ and d$_{[\overline{1}10]}$.}
\label{fig:Fig9}
\end{SCfigure*}

Fig.~\ref{fig:Fig7} plots the lowest three conduction band energies (e1, e2, e3) and the highest five valence band energies (h1, h2, h3, h4, h5) as a function of the elongation factor ($\eta$) for all of the three types of elongations: (a, d) Type-I, (b, e) Type-IIv, and (c, f) Type-II. In order to develop an understanding of these energy shifts, we also plot the wave functions and the piezoelectric potentials for a few selected values of $\eta$ in Figs.~\ref{fig:Fig8} and ~\ref{fig:Fig9}, respectively. 

The piezoelectric potential plots in Fig.~\ref{fig:Fig9} show familiar quadrupole symmetry, but exhibit three noticeable differences when compared to the plots for the flat QD (Fig.~\ref{fig:Fig4}): 

\begin{description}

\item [(i)] The interior of the QD is nearly field free, with only one peak at the QD interfaces. This is in agreement with Schliwa \textit{et al.}\cite{Schliwa_1} indicating a cancellation between the linear and the quadratic components inside the QD region.
\item [(ii)] A much larger magnitude of the fields (nearly twice).
\item [(iii)] A flip in the sign of the fields. The potential peaks are positive along the [$\overline{1}$10] direction and negative along the [110] direction. 

\end{description}

It is interesting to note here that although the piezoelectric potential profiles are drastically different for the flat and the tall QDs, but the electron and hole states for the circular-base case ($\eta$ = 1.0) align in the same direction for both cases: e2 along the [$\overline{1}$10] direction and all hole states along the [110] direction, as shown in the Fig.~\ref{fig:Fig8}. Thus, an experimental measurement on a circular-base dome-shaped QD would not indicate any sign difference for the e$_{[110]}$-e$_{[\overline{1}10]}$ as a function of QD AR, despite the underlying physical details are quite different and require atomistic modeling with realistic simulation domains and physical parameters.

We also find that the wave functions of the hole states for the tall QD are localized close to the QD interfaces due to the presence of the heavy hole (HH) pockets in the valence band edges. In our previous study\cite{Usman_5}, we have shown that such interfacial localization of the hole wave functions for a pure InAs QD starts when the AR is increased above 0.25. Similar results were presented by Narvaez \textit{et al.}\cite{Narvaez_1} for the InAs QDs using pseudo-potential calculations, where they varied the QD AR from (5/25.2) 0.198 to (7.5/25.2) 0.298 and showed that the hole states tend to confine in the HH pockets at the QD interfaces for the tall QDs.

\vspace{1mm}  

\vspace{1mm}      

\vspace{1mm}  

\textit{\textbf{Lowest three conduction band states:}} As for the flat QD of the previous subsection, the hydrostatic strain component does not change as a function of $\eta$, thus the strain contribution in the conduction band energy shifts is negligible because they are only affected by the hydrostatic strain. The piezoelectric fields also only slightly change when the QD is elongated (Fig.~\ref{fig:Fig9}), so the contribution from the changes in the piezoelectric fields in the shifts of the electron energies is minor, with dominating contributions coming from $\bigtriangleup$V and $\bigtriangleup$d.   

The lowest conduction band state e1 has s-type symmetry and the impact of $\bigtriangleup$d on its energy is negligible. The Type-I elongation considerably reduces the QD volume (see Fig.~\ref{fig:Fig1}(f)), and therefore e1 energy increases. When the QD volume is unchanged as in the case of the Type-IIv elongation or is only slightly decreased as in the case of the Type-II elongation, e1 shows a very small change. The wave function plots for the e1 state indicate s-type symmetry with only slight elongation, mainly for the [$\overline{1}$10] oriented Type-IIv and Type-II elongations. 

The excited electron states, e2 and e3, are separated by $\approx$8.8 meV for the circular-base QD ($\eta$=1.0), which is around four times larger than the $\approx$2.4 meV splitting for the flat QD. This larger splitting is mainly due to the larger (nearly twice) magnitude of the piezoelectric potentials as evident from the comparison of Figs.~\ref{fig:Fig4} and ~\ref{fig:Fig9}.   

For the Type-I elongation (Fig.~\ref{fig:Fig7}(a)), the large reduction in the QD volume shifts both, e2 and e3, towards the higher energies. The energy difference $\bigtriangleup$e$_p$ = e3-e2 increases for the [$\overline{1}$10]-elongation and decreases for the [110]-elongation, which is in agreement with what Schliwa \textit{et al.}~\cite{Schliwa_1} also calculated for the square-base pyramidal-shaped QDs. They reported electron p-state degeneracy towards the [110] base elongation. 

When the overall QD volume is fixed in the [$\overline{1}$10] Type-IIv elongation, the $\bigtriangleup$d induced shift lowers the energy of the [$\overline{1}$10]-oriented e2 and increases the energy of the [110]-oriented e3, as shown in Fig.~\ref{fig:Fig7}(b). Since e2 and e3 go through a flip of their orientations for the [110] Type-IIv elongation, so only a very small change in their energies is observed (from 8.8 meV to 9.3 meV).  

Finally, the changes in the energies of e2 and e3 for the Type-II elongations (Fig.~\ref{fig:Fig7}(c)) follow the similar trends as previously calculated for the flat QD (Fig.~\ref{fig:Fig2}(c)). The two competing factors ($\bigtriangleup$V and $\bigtriangleup$d) contribute to the energy shifts, where $\bigtriangleup$d being dominant for the small elongations ($\eta$ = 0.67, 0.82, 1.22, and 1.5) and $\bigtriangleup$V taking charge for the large elongations ($\eta$ = 0.54 and 1.86). 

\textbf{\textit{Separation between the lowest two electron energies:}} The difference between the lowest two conduction band energy levels ($\bigtriangleup$e$_{21}$ = e2-e1) which is important for the laser design robustness increases for the Type-I elongations, with largest increase being $\approx$8 meV calculated for the [110]-elongation. For the Type-II and Type-IIv elongations, only minor reductions in the $\bigtriangleup$e$_{21}$ are calculated. Therefore, we extend our conclusion of the previous subsection that the elongations of both, the flat and the tall QDs, do not deteriorate $\bigtriangleup$e$_{21}$ for the implementation of laser operation. 

\textit{\textbf{Highest five valence band energies:}} While the conduction band energy level shifts for the tall QD quite resemble with the corresponding shifts for the flat QD, the hole energy level shifts show significant contrasts. The main reason for this different behavior is the dissimilarity of the net piezoelectric potential profiles for the two types of the QDs as evident from the comparison of Figs.~\ref{fig:Fig4} and ~\ref{fig:Fig9}, which leads to very different confinements of the hole wave functions (see Figs.~\ref{fig:Fig3} and ~\ref{fig:Fig8}). The large negative peaks of the potentials at the QD interfaces along the [110] direction and nearly field free interior of the tall QD result in the hole wave functions being oriented along the [110] direction irrespective of the type and the direction of the elongations as shown in Fig.~\ref{fig:Fig8}. Furthermore, the HH pockets at the QD interfaces results in the hole wave function confinements at the QD interfaces.

\begin{figure*}
\includegraphics[scale=0.3]{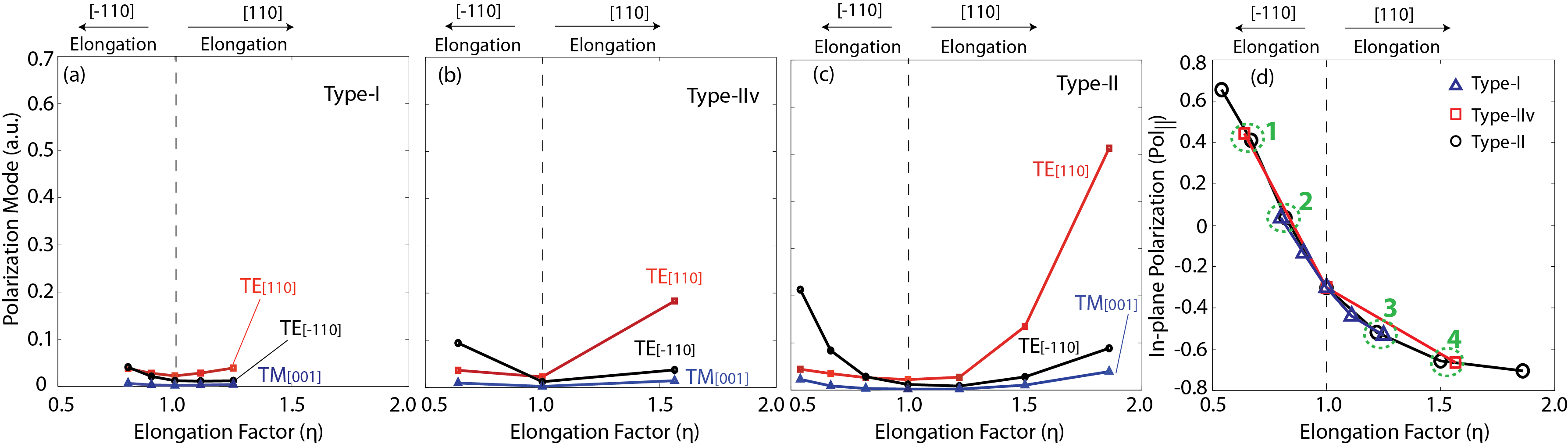}
\caption{The polarization dependent optical transition modes TE$_{[110]}$, TE$_{[\overline{1}10}]$, and TM$_{[001]}$ are drawn as a function of the (a) Type-I, (b) Type-IIv, and (c) Type-II QD elongations. The figures are plotted using the same scales to facilitate easy mutual comparison. (d) The plots of the in-plane polarization anisotropy (Pol$_{||}$) as defined by Eq.~\ref{eq:pol} are shown for the three type of the elongations, exhibiting an inverse quadratic dependence on $\eta$, consistent with the Sheng \textit{et al.}\cite{Sheng_1}. Four cases are also marked by using green ovals indicating that the two different types of elongations, with roughly similar values of $\eta$, exhibit similar values of the Pol$_{||}$.}
\label{fig:Fig10}
\end{figure*}
\vspace{1mm} 

For the Type-I elongation (Fig.~\ref{fig:Fig7}(d)), both a large decrease in the QD volume ($\bigtriangleup$V) and a small relaxation of the biaxial strain pushes the hole energy levels towards the lower energies. When the QD is elliptical along the [110] direction, the hole wave functions also being oriented along this direction experience a downward shift in their energies. Thus all of the three factors add to result in a stronger reduction in the hole energy levels. In the case of the [$\overline{1}$10] elongation, d$_{[110]}$ is reduced which pushes the hole energy levels upward. Although the cumulative downward shift from $\bigtriangleup$V and biaxial strain relaxation overall remains dominant, however for some values of $\eta$, $\bigtriangleup$d induced upward shift is evident.

As $\bigtriangleup$V=0 for the Type-IIv elongation and the biaxial strain relaxation is much weaker as compared to the Type-I case, so only $\bigtriangleup$d induced shifts are observed in Fig.~\ref{fig:Fig7}(e): the increase in d$_{[110]}$ decreases the hole energies (for the [110] elongation) and the decrease in d$_{[110]}$ increases the hole energies (for the [$\overline{1}$10] elongation). 

Finally, for the Type-II elongation in Fig.~\ref{fig:Fig7}(f), $\bigtriangleup$V is quite small. The biaxial strain relaxation again causes a small downward shift in the hole energies. However, the dominant shift comes from the $\bigtriangleup$d. Therefore, the hole energies move towards the lower values for the [110] elongation and shifts towards the higher values for the [$\overline{1}$10] elongation.

\textit{\textbf{Optical gap energy, E$_{g}$:}} The changes in the optical gap energy, E$_{g}$ = e1 - h1, are also marked with the help of vertical arrows (using red color) for the three types of the elongations in Fig.~\ref{fig:Fig7}. For the Type-I elongation, the increase in e1 energy and the decrease in h1 energy implies a blue shift of E$_{g}$ irrespective of the orientation of the elongation. This is same as earlier calculated for the flat QD. However, whereas for the flat QD, E$_{g}$ red shifts for both the Type-II and the Type-IIv elongations independent of their orientations, in this case of the tall QD, the changes in E$_{g}$ depend on the orientation of the elongation. For the [$\overline{1}$10] elongations, E$_{g}$ red shifts, whereas it blue shifts for the [110] oriented elongations.

\subsubsection{Polarization properties of the tall QD}

The polarization properties of the tall QD are significantly different from the flat QD of the previous subsection because of the two major differences in the hole wave functions even for the circular-base case, as evident from the comparison of Figs.~\ref{fig:Fig3} and ~\ref{fig:Fig8}: all of the hole wave functions for the tall QD are oriented along the [110] direction and are confined inside HH pockets at the QD interfaces in contrast to the flat QD where the hole wave functions are either nearly symmetric at the QD center or [$\overline{1}$10] oriented. Thus, for the circular-base tall QD, we calculate TE$_{[110]} <$ TE$_{[\overline{1}10]}$, and overall TE and TM modes have smaller magnitudes compared to the flat QD due to the relatively smaller spatial overlaps between the electron and the hole wave functions.

\begin{table*}
\caption{\label{tab:table2} The ratio of the LH components in the highest five valence band states (h1, h2, h3, h4, h5) of the elliptical-shaped tall QD with respect to the LH components of the corresponding valence band states of the circular-base ($\eta$) QD for a few selected values of the elongation factor, $\eta$. For each case, the corresponding ratio of the electron-hole wave function spatial overlap along the $\vec{z}$ = [001] direction given by e1hi = $| \langle \psi_{e1} | \vec{z} | \psi_{hi} \rangle |$ is also provided.}
\renewcommand{\arraystretch}{1.5}
\begin{tabular}{|l|l|l|l|l|l|l|}
\multicolumn{7}{c}{} \\
\cline{1-7}
\multicolumn{1}{|c|}{\textbf{Elongation type}} & 
\multicolumn{1}{c|}{\textbf{$\pmb{\eta}$}} & 
\multicolumn{1}{c|}{\textbf{h1 (e1h1)}} &
\multicolumn{1}{c|}{\textbf{h2 (e1h2)}} &
\multicolumn{1}{c|}{\textbf{h3 (e1h3)}} &
\multicolumn{1}{c|}{\textbf{h4 (e1h4)}} &
\multicolumn{1}{c|}{\textbf{h5 (e1h5)}} \\ 
\cline{1-7}
\multicolumn{1}{|c|}{\multirow{2}{*}{\textbf{Type-I}}} & 1.25 & 1.17 (1.42) & 1.18 (1.28) & 1.21 (1.40) & 1.21 (1.53) & 1.12 (1.25) \\ 
\cline{2-7}
\multicolumn{1}{|c|}{} & 0.80 & 1.07 (1.84) & 1.05 (1.89) & 1.04 (1.32) & 1.06 (1.50) & 1.00 (1.64) \\ 
\cline{1-7}
\multicolumn{1}{|c|}{\multirow{2}{*}{\textbf{Type-IIv}}} & 1.565 & 1.17 (1.45) & 1.15 (1.63) & 1.18 (1.73) & 1.15 (3.00) & 1.10 (3.89) \\ 
\cline{2-7}
\multicolumn{1}{|c|}{} & 0.64 & 0.93 (2.92) & 0.93 (3.34) & 0.85 (1.85) & 0.86 (2.14) & 0.86 (2.66) \\ 
\cline{1-7}
\multicolumn{1}{|c|}{\multirow{2}{*}{\textbf{Type-II}}} & 1.86 & 1.27 (3.29) & 1.26 (4.95) & 1.00 (5.80) & 1.08 (3.86) & 1.15 (4.39) \\ 
\cline{2-7}
\multicolumn{1}{|c|}{} & 0.54 & 1.00 (5.88) & 1.00 (6.17) & 0.86 (3.07) & 0.89 (3.19) & 0.90 (4.06) \\ 
\cline{1-7}
\end{tabular} 
\end{table*} 

\begin{figure*}
\includegraphics[scale=0.32]{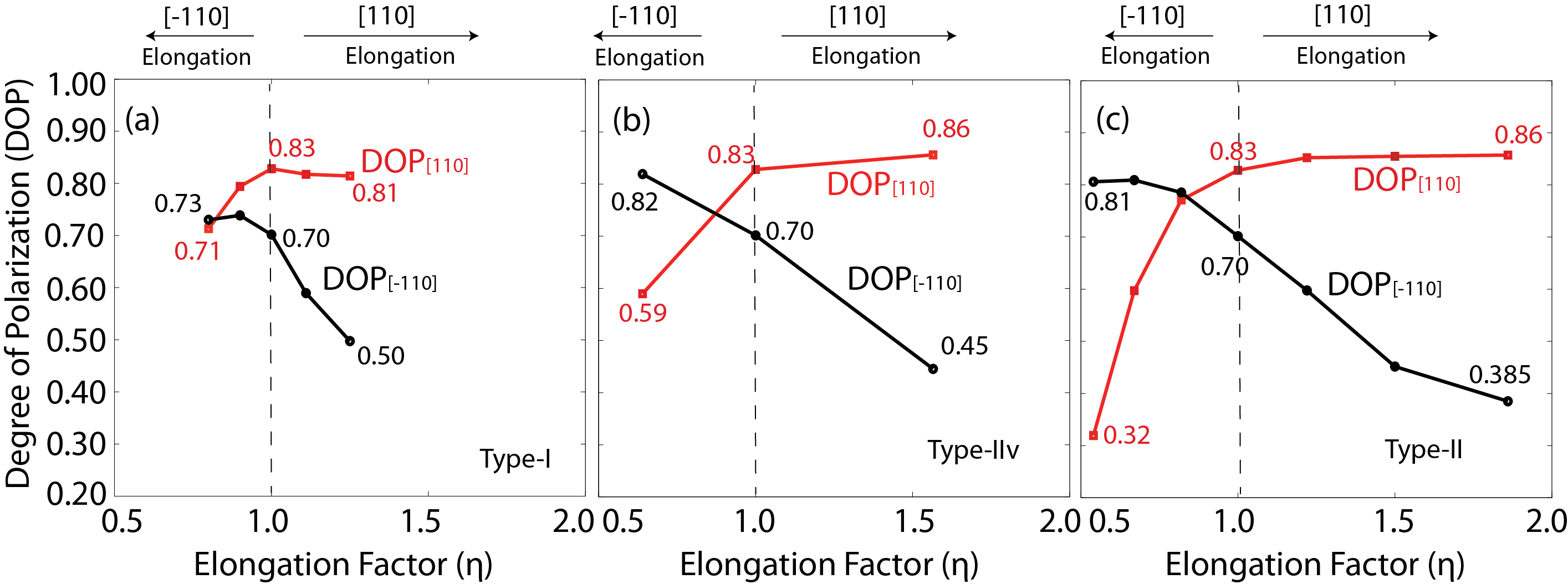}
\caption{The plots of the DOP$_{[110]}$ and DOP$_{[\overline{1}10]}$ are shown as a function of the (a) Type-I, (b) Type-IIv, and (c) Type-II elongations. A large decrease in the value of the DOP$_{[\vec{n}]}$ along the minor-axis of the elliptical QDs is calculated.}
\label{fig:Fig11}
\end{figure*}
\vspace{1mm}  

We also want to specify that for the tall QD, the lower lying hole wave functions have larger contribution in the ground state optical transition intensity when compared to the flat QDs where mainly e1-h1 transition is dominant\cite{Usman_5}. The dependence on the elongation factor is also quite different with all the polarization modes TE$_{[110]}$, TE$_{[\overline{1}10]}$, and TM$_{[001]}$ increasing irrespective of the type and the orientation of the elongation. The large increase in the TE$_{[\overrightarrow{n}]}$ mode for the elongation along the [$\overrightarrow{n}$] is same as also calculated for the flat QD and is explained in the section IV-A-2. However, the small increase in the other TE mode is attributed to an increased electron-hole spatial overlap because of elliptical shape of the QD.

Table~\ref{tab:table2} provides the ratio of the LH components in the top five valence band states for the elliptical QDs with respect to the circular-base QD. In each case, we also provide the corresponding ratio of the spatial overlap between the electron and hole wave functions (e1hi) along the [001]-direction as defined earlier for the flat QD. The increase in the TM$_{[001]}$ mode is directly related to an increase in the LH mixing of the valence band states. However for a giving LH mixing the magnitude of the TM$_{[001]}$ is proportional to the spatial overlap between the electron and hole wave functions along the [001] direction. 

As the shape of the QD is elongated, the electron wave function e1 also gets elongated along the major axis of the ellipse as evident from Fig.~\ref{fig:Fig8}. This increased spread of e1 wave function results in a relative increase in its spatial overlap with the hole wave functions for the elliptical QDs with respect to the circular-base QD as also noticeable from the values of e1hi provided in Table~\ref{tab:table2}. For $\eta >$ 1.0 ([110] elongation), the LH mixing increases and along with an increase in e1hi is responsible for the enhanced TM$_{[001]}$ mode. The [$\overline{1}$10] elongation ($\eta <$ 1.0) slightly reduces the LH character of the valence band states, however as the hole wave functions are oriented along the minor-axis, so a large increase in the electron-hole wave function spatial overlap overcomes a small ($<$ 15\%) decrease in the LH component and therefore causes a net increase in the TM$_{[001]}$ mode strength. It should also be noted that for the same amount of elongation, $\eta >$ 1.0 results in larger TM$_{[001]}$ mode when compared to $\eta <$ 1.0. This is because of the fact that for the $\eta >$ 1.0, both the LH mixing and the spatial overlap increase, whereas for the $\eta <$ 1.0 the LH mixing decreases and only the spatial overlap increases. 

\textbf{\textit{In-plane polarization anisotropy:}} As shown in Fig.~\ref{fig:Fig10}(d), the dependence of Pol$_{||}$ on the elongation factor ($\eta$) for the tall QD is also quite similar to the case of the flat QD. The calculated values of Pol$_{||}$ for all of the three types of the elongations again follow inverse quadratic dependence. Furthermore, for the fixed height of the QD, similar values of $\eta$ exhibit nearly same in-plane anisotropy. We highlight four such cases in Fig.~\ref{fig:Fig5}(d) using green dotted circles where nearly same values of the Pol$_{||}$ are calculated for the following sets of the values of $\eta$: (1) 0.64 in Type-IIv and 0.67 in Type-II, (2) 0.82 in Type-II and 0.8 in Type-I, (3) 1.22 in Type-II and 1.25 in Type-I, and (4) 1.565 in Type-IIv and 1.50 in Type-II. Therefore, we extend our conclusion of the previous subsection that the in-plane polarization anisotropy (Pol$_{||}$) only depends on the value of the $\eta$, irrespective of the type and orientation of the elongation for both flat and tall QDs, provided the height of the QDs is kept fixed.      

\textbf{\textit{Tuning of DOP$_{[\overrightarrow{n}]}$:}} To conclude this discussion about the polarization response of the tall QD, we compare the values of the DOP$_{[110]}$ and DOP$_{[\overline{1}10]}$ for the three types of elongations in Figs.~\ref{fig:Fig11}(a), (b), and (c). Since TE$_{[110]} >$ TE$_{[\overline{1}10]}$ for the circular-base ($\eta$ = 1.0), so a much larger difference ($\approx$13$\%$) is present between the DOP$_{[110]}$ and DOP$_{[\overline{1}10]}$ for $\eta$ = 1.0, as compared to only $\approx$0.2$\%$ difference for the flat QD. This difference is in agreement with the previous comparison\cite{Usman_5} between the similar flat and tall QDs and is attributed to the stronger orientation and confinements of the hole wave functions for the tall QDs. 

The elliptical shape of the tall QD once again reduces the value of the DOP$_{[\overrightarrow{n}]}$ along its minor-axis. This is mainly because the TM$_{[001]}$-mode increases for the elongated QDs whereas the corresponding TE-mode along the minor-axis does not increase much. In contrast to the case of the flat QD, the elliptical shape, however, does not reduce the value of the DOP$_{[\overrightarrow{n}]}$ along the major-axis which in fact first slightly increases and then remains nearly constant. This is because for the flat QD, the TE-mode is much larger than the TM-mode and hence the values of the DOP$_{[\overrightarrow{n}]}$ are not very sensitive to changes in the TM-mode. However, for the tall QD, the TE-modes are comparatively smaller in magnitude due to the smaller electron-hole spatial overlaps, so the values of the DOP$_{[\overrightarrow{n}]}$ become more sensitive to the changes in the TE- and TM-modes. For the small values of $\eta$, a larger increase in the TE-mode produces slight increase in the values of the DOP$_{[\overrightarrow{n}]}$. For the large values of $\eta$, the increase in the TM-mode also becomes important and hence the values of the DOP$_{[\overrightarrow{n}]}$ do not show any further increase. 

To summarize this subsection, we find that overall a much larger tuning of the polarization response is possible by elongating the tall QDs. The largest reduction ($\approx$51$\%$) in the value of the DOP$_{[\overrightarrow{n}]}$ is calculated for the Type-II elongation at $\eta$ = 0.54. Since the red shift of the optical wavelength and the isotropic polarization response, both, are desired for the design of the optical devices operating at telecommunication wavelengths (1.3-1.5 $\mu$m), our model calculations suggest that the Type-II [$\overline{1}$10] elongations are more suitable as they fulfil both requirements (see Figs.~\ref{fig:Fig7}(c), (f) and Fig.~\ref{fig:Fig11}(c)). It should also be noted that the confinements of the hole wave functions at the interfaces for the circular-base tall QDs reduce the oscillator strengths by an order of magnitude when compared to the circular-base flat QDs. However, with the elliptical shapes, the oscillator strengths increase and even become comparable to the flat QDs, in particular for the large Type-II elongations. 


\subsection{Vertical Stack of Nine QDs (9-VQDS)}

Vertical stacks of QDs (VSQDs) have shown great potential for tuning of the polarization properties. Recent experiments~\cite{Inoue_1, Alonso_1, Humlicek_1, Fortunato_1} and theoretical investigations~\cite{Usman_1, Saito_1} have demonstrated that an isotropic polarization response can be realized by geometrical engineering of the VSQDs. In this subsection, we study a vertical stack of closely spaced nine QD layers (9-VSQDs) as shown by the schematic diagram of Fig.~\ref{fig:Fig1}(c). This 9-VSQDs has been a topic of the recent studies ~\cite{Usman_1, Inoue_1, Saito_1} due to its significant technological relevance to achieve isotropic polarization for the implementation of the QD based SOA's. The optimized geometrical parameters of the 9-VSQDs are chosen directly from the experiment~\cite{Inoue_1}, so that our results remain relevant to the experimental community.

We have recently shown\cite{Usman_1} by experimental PL measurements and theoretical calculations that the 9-VSQDs can exhibit TM$_{[001]} >$ TE$_{[110]}$ leading to DOP$_{[110]} < 0$. However, a significant anisotropy in the in-plane TE-mode was measured resulting in TE$_{[\overline{1}10]} >$ TM$_{[001]}$ and DOP$_{[\overline{1}10]} >$ 0. Similar anisotropies in the DOP$_{[\overrightarrow{n}]}$ were independently measured by alonso-Alvarez \textit{et al.}~\cite{Alonso_1} and Humlicek \textit{et al.}~\cite{Humlicek_1}, which they were unable to explain. Our multi-million atom simulations~\cite{Usman_1} assuming circular-base for the QDs qualitatively explained that this anisotropy (DOP$_{[110]} \neq$ DOP$_{[\overline{1}10]}$ ) is due to a strong confinement of the hole wave functions at the interfaces of QDs (similar to the case of the tall QD of the previous subsection) which tend to align along the [$\overline{1}10]$-direction, and thus significantly reduce the TE$_{[110]}$-mode. The TE$_{[\overline{1}10]}$-mode, on the other hand, does not observe any such decrease. The small increase in the TM$_{[001]}$ mode due to the relaxation of the biaxial strain, in particular around the center of the 9-VSQDs\cite{Usman_1, Saito_1}, is also not sufficient to overcome the TE$_{[\overline{1}10]}$ mode and thus the DOP$_{[\overline{1}10]}$ remains considerably larger than zero.

\begin{figure*}
\includegraphics[scale=0.3]{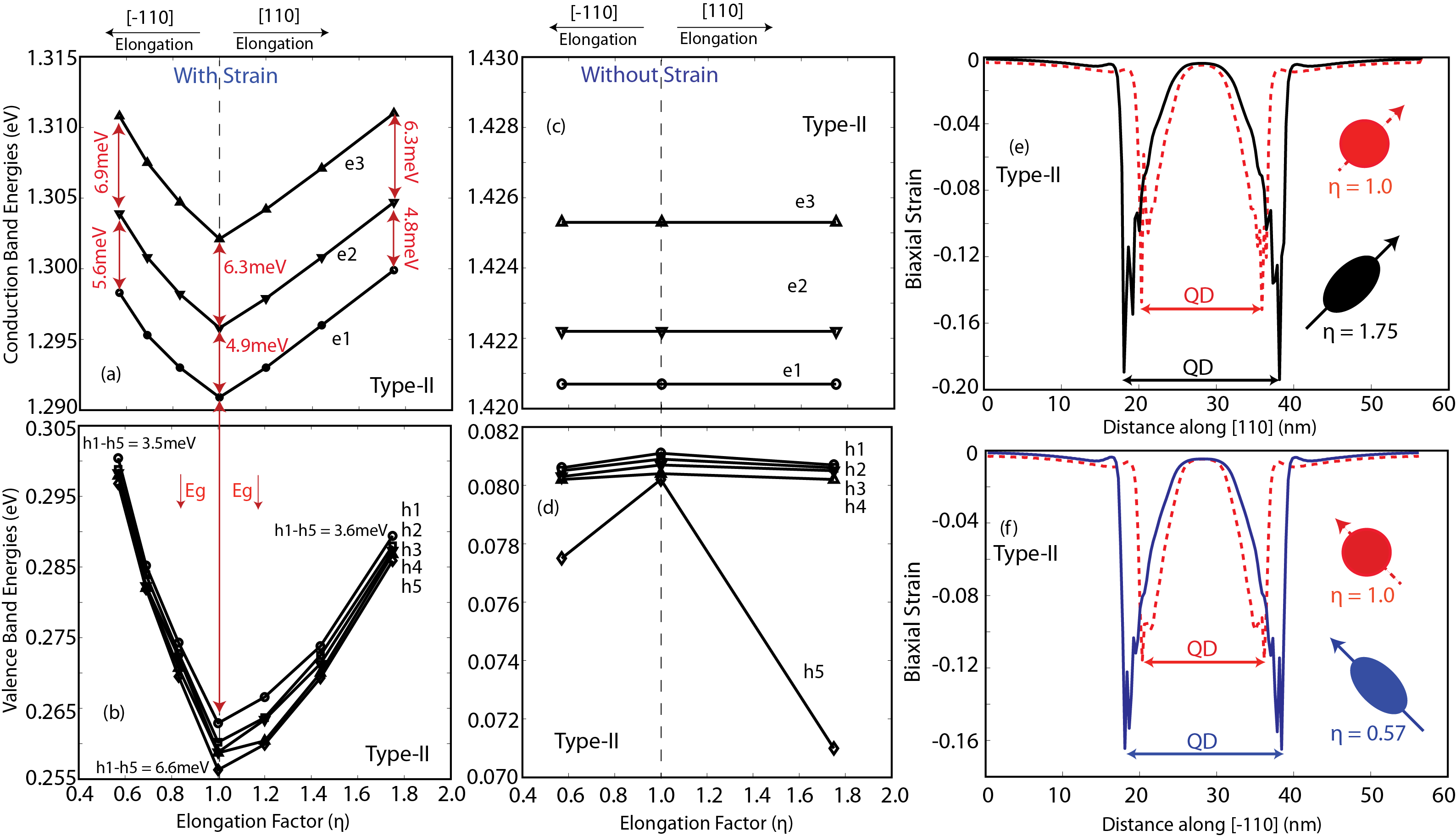}
\caption{The plots of (a) the lowest three conduction band energies (e1, e2, e3) and (b) the highest five valence band energies (h1, h2, h3, h4, h5) for the 9-VSQDs as a function of the Type-II elongation factor, $\eta$. All of the energies increase irrespective of the orientation of the elongation. (c, d) The plots of the electron and hole energies as in (a) and (b), but without including the effect of strain. In the absence of strain, only $\bigtriangleup$d and $\bigtriangleup$V contribute in the energy shifts. (e) The plots of the biaxial strain component ($\epsilon_{xx}+\epsilon_{yy}-2\epsilon_{zz}$) along the [110] direction through the center of the 9-VSQDs with circular-base ($\eta$=1) and the [110] elongated base ($\eta$=1.75). (f) The plots of the biaxial strain component ($\epsilon_{xx}+\epsilon_{yy}-2\epsilon_{zz}$) along the [$\overline{1}$10] direction through the center of the 9-VSQDs with circular-base ($\eta$=1) and the [$\overline{1}$10] elongated base ($\eta$=0.57).}
\label{fig:Fig12}
\end{figure*}
\vspace{1mm} 

The good agreement of our theoretical results with the experimental PL measurements even for an ideal circular-base 9-VSQDs shape leads to a fundamental question that how much is the contribution from the realistic shapes which are normally elongated. Here we systematically elongate the QD layers inside the 9-VSQDs base and analyse its impact on the electronic and polarization properties. In contrast to the single QDs of the previous two subsections, where the QD base elongations result in a significant tuning of the DOP$_{[\overrightarrow{n}]}$, the magnitude of the DOP$_{[\overrightarrow{n}]}$ for the 9-VSQDs is relatively insensitive to the value of $\eta$. However, the sign of the DOP$_{[\overrightarrow{n}]}$ is a strong function of the orientation of the elongation, and even a very small elongation (0.5-1.0 nm) is sufficient to control the sign of the DOP$_{[\overrightarrow{n}]}$. Furthermore, we explore the possibility to achieve DOP$_{[\overrightarrow{n}]} <$ 0 for both $[\overrightarrow{n}]$ = [110] and [$\overline{1}$10] by elongating the 9-VSQDs along the [110] direction. Our calculations predict that such a scenario is not possible due to a very high sensitivity of the DOP$_{[\overrightarrow{n}]}$ with the value of $\eta$, and therefore the value of the DOP$_{[\overrightarrow{n}]}$ may be reduced below zero for only one of the two spatial directions.  

\begin{figure*}
\includegraphics[scale=0.25]{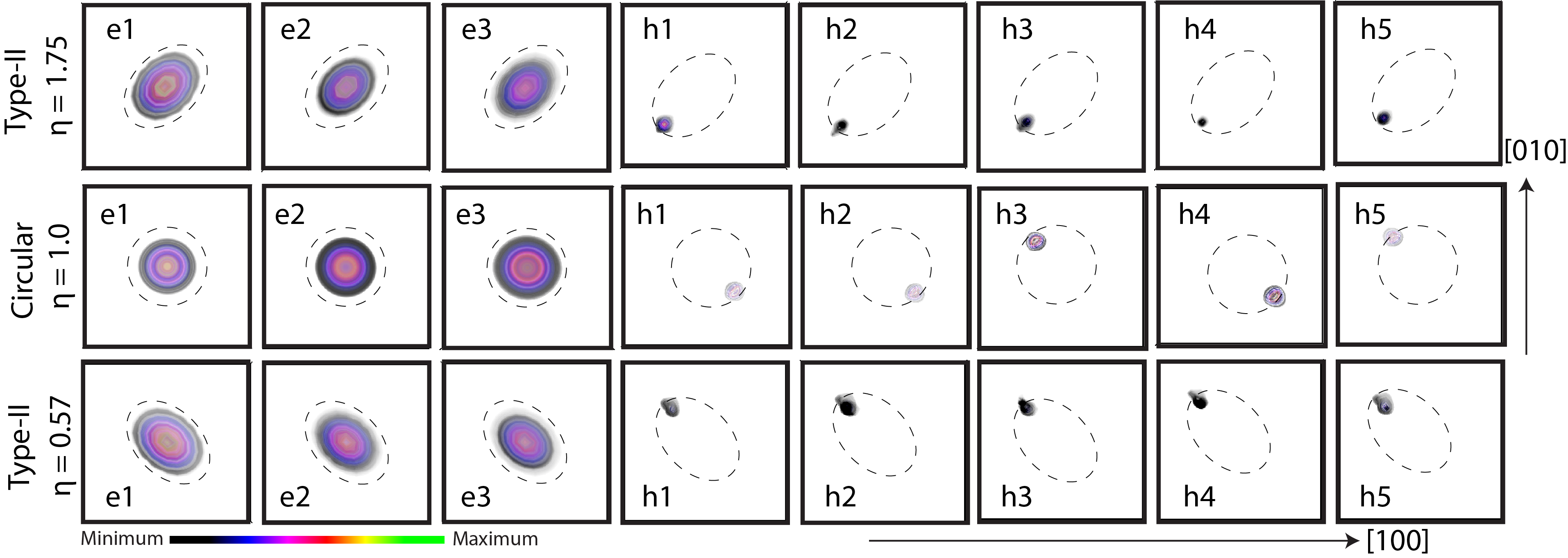}
\caption{Top views of the wave function plots for the lowest three conduction band (e1, e2, and e3) and the highest five valence band (h1, h2, h3, h4, and h5) states are shown for the circular-base 9-VSQDs and the two selected Type-II elongations of the 9-VSQD. The intensity of the colors in the plots represent the magnitude of the wave functions, with the dark color color indicating the smallest magnitude and the light green color indicating the largest magnitude. The boundaries of the QDs are also shown to guide the eye. }
\label{fig:Fig13}
\end{figure*}
\vspace{1mm}  

\subsubsection{Electronic properties of the 9-VSQDs}             

\textbf{\textit{Lowest three conduction band energies:}} Fig.~\ref{fig:Fig12}(a) plots the lowest three conduction band energies (e1, e2, e3) as a function of the elongation factor ($\eta$) for the Type-II elongation. The corresponding wave functions are also shown in Fig.~\ref{fig:Fig13}, indicating that all of the lowest three conduction band states have s-type symmetry. Due to the strong coupling between the closely spaced QD layers, these electron states are hybridized over multiple QD layers\cite{Usman_1}. Only a very small increase (less than 10 meV) in the energies of e1, e2, and e3 is calculated as the 9-VSQDs is elongated. Due to the s-type symmetry, the effect of the $\bigtriangleup$d is negligible. We also find that in contrast to the single QDs, the energies of these molecular electron states are insensitive to the $\bigtriangleup$V, as also confirmed by Fig.~\ref{fig:Fig12}(c) where no shift in the e1, e2, and e3 energies is calculated when the impact of strain is excluded. This is due to the fact that the electron energies are spread over multiple QD layers as well as occupies the GaAs spacers in between them, so a small decrease in the QD volume due to the Type-II elongation only negligibly impact their energies. Our calculations find that the hydrostatic component of the strain ($\epsilon_{xx}+\epsilon_{yy}+\epsilon_{zz}$) slightly increases for the elongated 9-VSQDs and results in a small upward shift in the conduction band energies. Also, the shifts in the energies are nearly equal for e1, e2, and e3, and therefore the separation between them remains nearly unchanged.

\begin{figure*}
\includegraphics[scale=0.45]{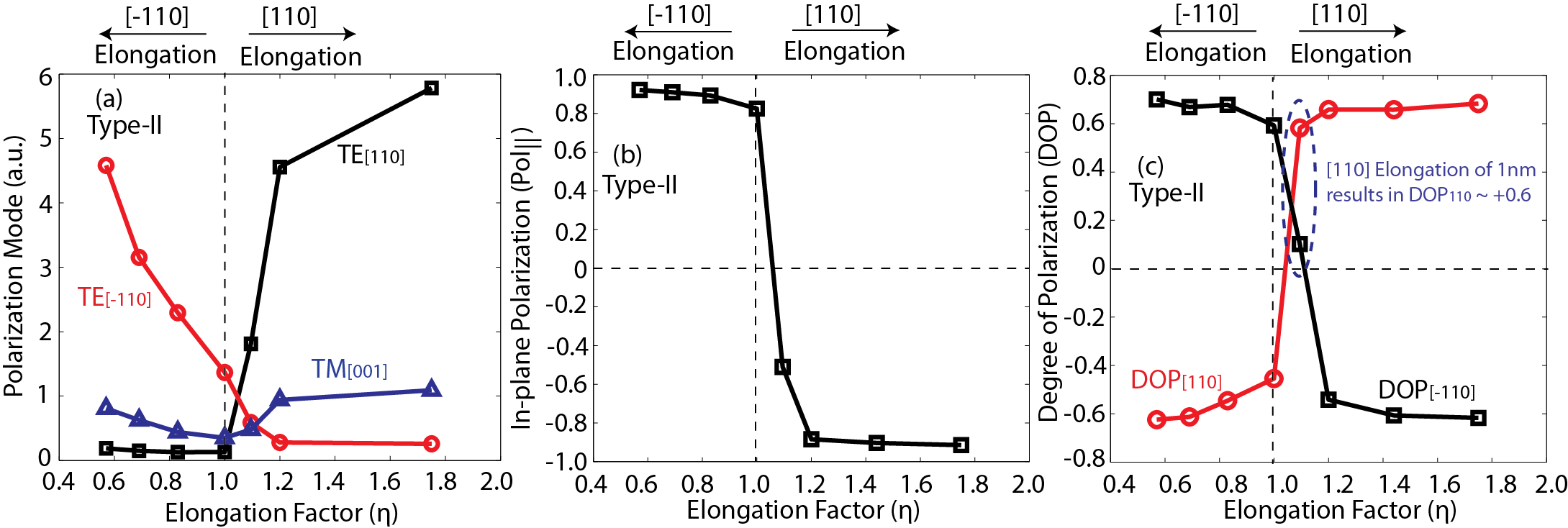}
\caption{(a) The plots of the polarization dependent TE and TM modes as a function of the Type-II elongation factor $\eta$ for the 9-VSQDs. (b) Plot of the POl$_{||}$ as a function of the Type-II elongation factor for the 9-VSQDs. The plots indicate a high degree of in-plane anisotropy for the 9-VSQDs, which is very sensitive to the elongation factor. (c) Plots of the DOP$_{[110]}$ and DOP$_{[\overline{1}10]}$ as a function of the Type-II elongation factor for the 9-VSQDs. For the circular-base 9-VSQDs, $\eta$=1.0, the DOP$_{[110]}$ is negative and the DOP$_{[\overline{1}10]}$ is positive. As we [110]-elongate the 9-VSQDs by 1 nm (marked by blue oval), the DOP$_{[\overline{1}10]}$ becomes close to zero, however the DOP$_{[110]}$ drastically increases to +0.60, suggesting that it is not possible to simultaneously engineer both DOPs below or close to zero for this 9-VSQDs using [110] elongation engineering.}
\label{fig:Fig14}
\end{figure*}
\vspace{1mm}

\textbf{\textit{Highest five valence band energies:}} Fig.~\ref{fig:Fig12}(b) plots the energies for the highest five valence band states (h1, h2, h3, h4, and h5) as a function of the Type-II elongation. All of the hole energies increase for both, the [110] and the [$\overline{1}10$], elongations. Fig~\ref{fig:Fig13} shows the top view of the hole wave functions for $\eta$=0.57, 1.0, and 1.75. For the circular-base 9-VSQDs, all the hole wave functions are aligned along the [$\overline{1}$10] direction due to the presence of the HH pockets, similar to the case of the single QDs with large aspect ratios\cite{Usman_5, Narvaez_1} and the bilayer QDs.\cite{Usman_2} 

For the elliptical 9-VSQDs, the hole wave functions align along the major-axis. This is due to the larger biaxial strain along these directions which pushes the HH pockets towards higher energies and hence the hole wave functions residing in these pockets also move towards higher energies. Fig.~\ref{fig:Fig12}(e) and (f) plots the biaxial strain components ($\epsilon_{xx}+\epsilon_{yy}-2\epsilon_{zz}$) through the center of the 9-VSQDs along the [110] and the [$\overline{1}$10] directions, respectively. The large increase in the biaxial strain is clearly evident for the elliptical 9-VSQDs when compared to the circular-base case. Since the hole energies are pushed up by an increase in the biaxial strain component, this shift dominates the small downward shift due to a small increase in the hydrostatic component ($\epsilon_{xx}+\epsilon_{yy}+\epsilon_{zz}$). As a result, all of the hole energies are shifted towards the higher values as evident from the Fig.~\ref{fig:Fig12}(b). The contributions from $\bigtriangleup$d and $\bigtriangleup$V are also very small, which is confirmed from Fig.~\ref{fig:Fig12}(d), where the strain effect is excluded, and the hole energies move towards lower values due to the combined shift induced by $\bigtriangleup$d and $\bigtriangleup$V. 

In Fig.~\ref{fig:Fig12}(b), a relatively smaller increase in the hole energies for the [110] elongation as compared to the [$\overline{1}$10] elongations is due to the fact that all of the hole wave functions are initially oriented along the [$\overline{1}$10] direction for the circular-base ($\eta$ = 1.0), and they go through a 90$^\circ$ rotation to align along the [110] direction for the [110] elongation. It should also be noted that the hole energy separations reduce as a function of the elongation factor ($\eta$) for the 9-VSQDs, suggesting even enhanced contributions from the lower lying valence band states in the ground state optical intensity measured at the room temperature for the elliptical 9-VSQDs.

\begin{figure*}
\includegraphics[scale=0.33]{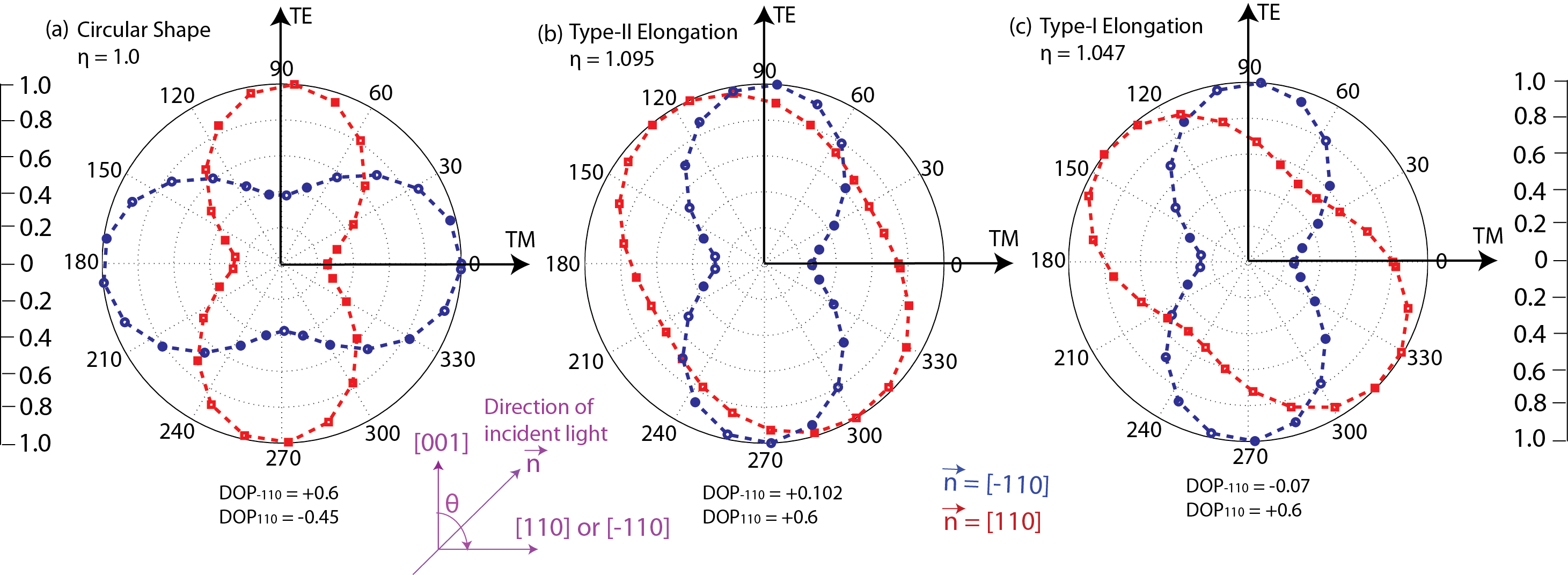}
\caption{Normalized Polar plots for the 9-VSQDs system with (a) circular-base, (b) Type-II [110]-elongation ($\eta$=1.095), and (c) Type-I [110]-elongation ($\eta$=1.047). In each case, the polarization direction
of the incident light (n) is kept along the [110]-direction (red squares) for the TE$_{[\overline{1}10]}$-mode and along the [$\overline{1}$10]-direction (blue circles) for the TE$_{[110]}$-mode. The polar plots based on the cumulative sum of the optical transition strengths between the lowest conduction band state and the highest five valence band states are drawn with respect to the angle $\theta$ between the [001]-direction and either the [110] or the [$\overline{1}$10] directions.}
\label{fig:Fig15}
\end{figure*}
\vspace{1mm}  

\textbf{\textit{Optical gap energy, E$_{g}$:}} From Figs.~\ref{fig:Fig12}(a) and (b), both the electron and hole energies increase as a function of the elongation, however the hole energies have larger slope and therefore the optical gap energy, E$_{g}$ = e$_{1}$ - h$_{1}$, decreases as a function of the elongation factor ($\eta$). We calculate a maximum red shift of $\approx$30 meV in E$_g$ for $\eta$ = 0.57. 

\subsubsection{Polarization properties of the 9-VSQDs}

Fig.~\ref{fig:Fig14}(a) plots the polarization dependent TE$_{[110]}$, TE$_{[\overline{1}10]}$, and TM$_{[001]}$-modes as a function of the Type-II elongation. For the circular-base ($\eta$=1.0) case, TE$_{[110]} <$ TM$_{[001]}$ and TE$_{[\overline{1}10]} >$ TE$_{[110]}$.

When the base of the 9-VSQDs is elongated, the increase in the biaxial strain component (see Figs.~\ref{fig:Fig12}(e) and (f)) increases the splitting between the LH and the HH bands, which will decrease the TM$_{[001]}$-mode. However, the small increase in the TM$_{[001]}$-mode as plotted in the Fig.~\ref{fig:Fig14}(a), is due to the hole wave functions confining towards the middle of the 9-VSQDs for the elliptical shapes where the HH/LH intermixing is larger as compared to its edges\cite{Usman_1}. For example, by comparing the $\eta$=1.0 and $\eta$=0.57 cases, we find following shifts in the spatial positions of the top-most five hole wave functions within the 9-VSQDs: h1 moves from the QD layer 2 to the QD layer 3, h2 moves from the QD layer 3 to the QD layer 5, h3 moves from the QD layer 2 to the QD layer 4, h4 stays in the QD layer 8, and h5 moves from the QD layer 3 to the QD layer 5; here the QD layers in the 9-VSQDs are numbered from 1 to 9 starting from the bottom towards the top, as mentioned in the schematic diagram of Fig.~\ref{fig:Fig1}(c). Similar trends in the spatial confinements of the hole wave functions are observed for other values of $\eta$ which are responsible for the small increase in the TM$_{[001]}$-mode as a function of $\eta$.   

The magnitude of the TE$_{[\overrightarrow{n}]}$ modes is calculated to be very sensitive to the elongation of the 9-VSQDs. Even for a very small [110] elongations ($\leq$ 1 nm), the TE$_{[110]}$ mode quickly increases above the TM$_{[001]}$-mode. For example, for $\eta$=1.095, the TE$_{[\overline{1}10]}$ reduces to become close to the TM$_{[001]}$, but the TE$_{[110]}$ has already increased by a factor of $\approx$13. 

It should also be noted that while the elliptical shape increases the TE$_{[\overrightarrow{n}]}$ mode along its major-axis, it has only very little impact on the TE-mode along its minor-axis. When the 9-VSQDs is elongated along the [$\overline{1}$10] direction, the TE$_{[\overline{1}10]}$ increases, but the TE$_{[110]}$ mode remains nearly unchanged. Similarly for the [110] elongation, the TE$_{[\overline{1}10]}$-mode quickly decreases and then remains nearly unchanged for $\eta >$ 1.2.  

\textbf{\textit{In-plane polarization anisotropy:}} Fig.~\ref{fig:Fig14}(b) plots the in-plane polarization (Pol$_{||}$) defined by Eq.~\ref{eq:pol} as a function of the Type-II elongation factor ($\eta$). The large magnitudes of the Pol$_{||}$ for both [110] and [$\overline{1}10$] elongations suggest a high degree of the in-plane polarization anisotropy for the 9-VSQDs. Even for the perfectly circular-base ($\eta$=1.0), Pol$_{||}$ is $\approx$0.82, indicating that TE$_{[\overline{1}10]} \gg$ TE$_{[110]}$. Any elongation along the [$\overline{1}10$]-direction further increases this anisotropy. The [110]-elongation sharply increases TE$_{[110]}$-mode and changes the sign of Pol$_{||}$. This is because of the 90$^\circ$ rotation of the hole wave functions (see Fig.~\ref{fig:Fig13}). Even an elongation as small as of 1 nm along the [110]-direction can change the value of the Pol$_{||}$ from +0.82 to -0.51. Therefore, we conclude that the 9-VSQDs exhibits highly anisotropic in-plane polarizations. Similar in-plane polarization anisotropies were measured by Alonso-Alvarez \textit{et al.}\cite{Alonso_1} and Humlicek \textit{et al.}\cite{Humlicek_1} for the vertical QD stacks. 

It should also be noted that whereas the in-plane polarization (Pol$_{||}$) for the single QDs, irrespective of their AR, exhibits an inverse quadratic relation with respect to $\eta$ (see Figs.~\ref{fig:Fig5}(d) and ~\ref{fig:Fig10}(d)), it demonstrates nearly a step function like dependence on $\eta$ for the strongly coupled 9-VSQDs.    

\textbf{\textit{Tuning of DOP$_{[\overrightarrow{n}]}$:}} Fig.~\ref{fig:Fig14}(c) plots the DOP$_{[110]}$ and DOP$_{[\overline{1}10]}$ as a function of the Type-II elongation. For the ideal circular-base case ($\eta$ = 1.0), DOP$_{[110]}$ and DOP$_{[\overline{1}10]}$ have values of -0.45 and +0.6. This is in qualitative agreement with the experimental PL measurements\citep{Inoue_1} and leads to a question that how much impact would be from a realistic elliptical shape. Our calculations show that the orientation of the base elongation determines the sign of the DOP$_{[\overrightarrow{n}]}$, whereas the magnitude of the elongation (value of $\eta$) has a very little impact on the magnitude of the DOP$_{[\overrightarrow{n}]}$. This is clearly evident from Fig.~\ref{fig:Fig14}(c) for $\eta \leq$ 1.0 and for $\eta \geq$ 1.2, where a very small change in the magnitude of the DOP$_{[\overrightarrow{n}]}$ is observed as the value of $\eta$ is changed.   

The strong dependence of the sign of the DOP$_{[\overrightarrow{n}]}$ on the orientation of the elongation is highlighted by using an oval in the Fig.~\ref{fig:Fig14}(c), where even for a 1 nm [110]-elongation, the DOP$_{[110]}$ drastically changes its sign from -0.45 to +0.6. This large change in the value of DOP$_{[\overrightarrow{n}]}$ for 1.0 $< \eta <$ 1.2 is remarkable as it indicated that only a very small shape asymmetry is capable of overcoming the effect of atomistic symmtery lowering effect. This also implies that the elliptical shape of the 9-VSQDs can not be exploited to simultaneously engineer both DOP$_{[110]}$ and DOP$_{[\overline{1}10]}$ below zero.

We want to highlight that although a tuning of the DOP$_{[\overrightarrow{n}]}$ over a wide range of values is possible by the elongation of the single QDs, it remains relatively insensitive with respect to the magnitude of $\eta$ for the 9-VSQDs. Therefore, we expect that the elongation of the 9-VSQDs would not offer much improvement in its polarization response.              

\textit{\textbf{Polar plots:}} The strong impact of the [110]-elongations on the polarization properties of the 9-VSQDs is further confirmed in Fig.~\ref{fig:Fig15} by comparing the normalized polar plots for (a) circular-base ($\eta$=1.0), (b) Type-II 1 nm [110]-elongation ($\eta$=1.095), and (c) Type-I 1 nm [110]-elongation ($\eta$=1.047). In each case, two polar plots are drawn: (i) as a function of the angle $\theta$ between the [001]-direction and the [110]-direction for the TE$_{[110]}$ (blue circles); (ii) as a function of the angle $\theta$ between the [001]-direction and the [$\overline{1}10$]-direction for the TE$_{[\overline{1}10]}$ (red squares). 

As the 9-VSQDs is elongated, a 90$^\circ$ rotation of the polar plots is calculated for the TE$_{[110]}$-modes in both the Type-I and the Type-II cases. This clearly suggests that both [110] elongations will result in TE$_{[110]}$-mode $>$ TM$_{[001]}$-mode. For TE$_{[\overline{1}10]}$-mode, the polar plot rotates anti-clockwise by $\approx$30$^\circ$ and $\approx$45$^\circ$ for the Type-II and the Type-I elongations, respectively. This causes a reduction between the relative magnitudes of the TE$_{[\overline{1}10]}$-mode and the TM$_{[001]}$-modes, thus reducing DOP$_{[\overline{1}10]}$ from +0.6 to 0.102 in (b) and to -0.07 in (c).

\textit{\textbf{Geometry of the experimentally grown 9-VSQDs:}} The above discussion about the strong dependence of the polarization properties on the [110]-elongations allows us to theoretically probe the geometrical shape of the 9-VSQDs as grown by Inoue \textit{et al.}\cite{Inoue_1}. It was reported that the 9-VSQDs are not isotropic and the TEM images suggested very little anisotropy in the lateral extent\cite{Ikeuchi_1}, possibly a [$\overline{1}10$]-elongation~\cite{Kita_1}. Our multi-million-atom calculations show that the polarization response is very sensitive to the elongation factor ($\eta$) and even a 0.5-1.0 nm [110]-elongation increases DOP$_{[110]}$ above zero. Therefore according to our model results, the experimentally measured DOP$_{[110]}$ = -0.6 implies that the shape of the 9-VSQDs can only have [$\overline{1}$10] elongated base which confirms the findings from the TEM images~\cite{Kita_1}. It should also be noted that as the 9-VSQDs studied here has pure InAs QD layers, so our finding does not contradict with the conclusions of Mlinar \textit{et al.}\cite{Mlinar_1} where they report that for the alloyed InGaAs QDs, the alloy random configurations may significantly impact the polarization properties and make the correlation between the measured polarization response and the QD geometry unreliable.

\section{Summary and Conclusions}

We have performed multi-million-atom simulations to understand the impact of the elliptical shapes on the electronic and polarization properties of the single and the multi-layer vertical stacks of InAs QDs. The comparison between a flat QD and a tall QD, having aspect ratios of 0.225 and 0.40 respectively, reveals drastically different electronic and polarization properties as a function of their base elongation. The key outcomes of the comparison are: 

\begin{description}

\item [(i)] The quadratic component of the piezoelectric potential completely cancel the linear component inside the tall QD region, whereas only partial cancellation occurs for the flat QD. 
\item [(ii)] Although the stain and the piezoelectric potentials are drastically different for the flat and tall QDs, the lower electron p-state (e2) is oriented along the [$\overline{1}$10] direction for both systems.  
\item [(iii)] The hole wave functions are confined inside the flat QD mainly at its center, whereas they are confined at the QD interfaces inside the HH pockets for the tall QD. This leads to a reduction of the oscillator strengths for the circular-base tall QD by approximately an order of the magnitude due to smaller electron-hole wave function spatial overlaps. For the elliptical-shaped tall QDs, the oscillator strengths increase and even become comparable to the flat QD for some values of the elongation factor ($\eta$).  
\item [(iv)] The Type-I elongation, irrespective of its orientation, blue shifts the optical gap energy (E$_{g}$) for both, the flat and the tall QDs. The Type-II and Type-IIv elongations red shift E$_{g}$ for the flat QD irrespective of the elongation direction, whereas for the tall QD the shift in E$_{g}$ strongly depends on the orientation of the elongation: red shift of E$_{g}$ for the [$\overline{1}$10] elongation and blue shift for the [110] elongation. 
\item [(v)] The elliptical shape of the flat QD always improves its polarization properties by reducing the value of the DOP$_{[\overrightarrow{n}]}$, whereas it only reduces the DOP$_{[\overrightarrow{n}]}$ along the minor-axis for the tall QD. 
\item [(vi)] The elliptical shape of the tall QD allows a tuning of the DOP$_{[\overrightarrow{n}]}$ over a much wider range, when compared to the flat QD. This property can be further exploited in large stacks of strongly coupled QD layers, where essentially very high values of the ARs can be achieved. 

\end{description}

Although the understanding of the single layers of the QDs provides significant physical insight of the impact of the shape asymmetry, they do not lead to isotropic polarization response (DOP$_{[\overrightarrow{n}]} \sim$ 0) for the QD base elongations of up to 6 nm studied in this paper. Therefore, we prbextend our study of the elliptical shapes to the experimentally reported vertical stack of nine QDs (9-VSQDs) which has demonstrated DOP$_{[110]}$ $<$ 0. The key features of our analysis about the elliptical 9-VSQDs are:

\begin{description}

\item [(i)] In contrast to the single QD layers where the elliptical shape only very slightly reduces the magnitude of the biaxial stain, a significant increase in the magnitude of the biaxial strain is calculated for the 9-VSQDs. Therefore, the shifts in the hole energies are dominated by the changes in the biaxial strain, rather than $\bigtriangleup$V and $\bigtriangleup$d. 
\item [(ii)] The hole wave functions are confined inside the HH pockets at the interfaces of the 9-VSQDs, similar to the case of the tall QD. This introduces a large in-plane polarization anisotropy.
\item [(iii)] While the elongations of the single QDs result in the tuning of DOP$_{[\overrightarrow{n}]}$ over a wide range, the magnitude of DOP$_{[\overrightarrow{n}]}$ is largely insensitive to the magnitude of $\eta$ for the 9-VSQDs. Therefore, we conclude that the elliptical shapes of the 9-VSQDs do not provide any noticeable improvement in the polarization response. This is clearly evident from Fig.~\ref{fig:Fig14}(c) for $\eta \leq$ 1.0 and for $\eta \geq$ 1.2. 
\item [(iv)] Our calculations show that the sign of DOP$_{[\overrightarrow{n}]}$ is very sensitive to the orientation of the elongation. Even a very small variation of $\eta$ from 1.0 is capable of controlling the sign of the DOP$_{[\overrightarrow{n}]}$. We find that the in-plane polarization (Pol$_{||}$) roughly follows a step function like abrupt dependence on $\eta$, as compared to an inverse quadratic dependence for the single QDs. Such a large in-plane anisotropy of the polarization allows to accurately predict the shape elongation of the 9-VSQDs studied in this paper, in agreement with the TEM findings.  
\end{description}

In summary, we have presented a detailed analysis of the dependence of the polarization properties as a function of the elongation factor $\eta$ that would serve as a guidance to engineer the geometry parameters for the tuning of DOP$_{[\overrightarrow{n}]}$ from semiconductor QDs, which is a critical design parameter for several challenging applications. 


\textbf{\textit{Acknowledgements:}} The author gratefully acknowledges Stefan Schulz (Tyndall National Institute) for critically reading the manuscript and providing valuable suggestions. Computational resources are acknowledged from National Science Foundation (NSF) funded Network for Computational Nanotechnology (NCN) through \url{http://nanohub.org}. NEMO 3D simulator was developed in parts at NASA JPL/Caltech and Purdue University by a number of researchers supervised by Prof. Gerhard Klimeck (Purdue University), whom work have been cited in the corresponding references\cite{Klimeck_1, Klimeck_2, Klimeck_3}. NEMO 3D based open source tools are available at: \url{https://nanohub.org/groups/nemo_3d_distribution}. 


\bibliographystyle{apsrev4-1}

%

\end{document}